\pdfoutput=1

\documentclass{article}

\usepackage{arxiv}

\usepackage[utf8]{inputenc} 
\usepackage[T1]{fontenc}    
\usepackage{hyperref}       
\usepackage{url}            
\usepackage{booktabs}       
\usepackage{amsfonts}       
\usepackage{nicefrac}       
\usepackage{microtype}      
\usepackage{lipsum}		
\usepackage{graphicx}

\usepackage{subfig}  

\usepackage{amsmath} 

\usepackage{booktabs} 
\usepackage{etoolbox}
\usepackage{xspace}
\usepackage{datetime}
\usepackage{tabularx}
\usepackage{fancyvrb}
\usepackage{enumitem}
\usepackage[ruled,vlined,linesnumbered]{algorithm2e}
\usepackage{letltxmacro}

\renewcommand{\deg}{d}  

\newcommand{\din}{d_{in}}  

\newcommand{\nodes}{|\nodeset|}  

\newcommand{\edgeset}{E}  
\newcommand{\nodeset}{V}  

\newcommand{\G}{G}  

\newcommand{\m}{|\edgeset|}  

\DeclareMathOperator*{\argmax}{argmax} 

\newcommand{\grp}{g}  
\newcommand{\g}{\grp}  

\newcommand{\gsize}{|g|}  

\newcommand{\p}{p}  

\newcommand{\pe}{p_{edge}}  

\newcommand{\pn}{p_{node}}  

\newcommand{\q}{q}  

\renewcommand{\i}{\mathcal{I}}  

\newcommand{\qi}{Modularity}  
\newcommand{\modularityr}{\qi}  

\newcommand{\Qn}{Q_{node}}  

\newcommand{\U}{U}  

\newcommand{\uscore}{u}  
\newcommand{\lscore}{l}  

\newcommand{\bsn}{b_{node}}  
\newcommand{\bse}{b_{edge}}  

\newcommand{\bff}{}  

\newcommand{\avgPR}{avgPR~} 
\newcommand{\topPR}{topPR}

\newcommand{\mysim}{\raise.17ex\hbox{$\scriptstyle\sim$}}
\newcommand{\mytilde}{\raise.17ex\hbox{$\scriptstyle\sim$}}

\newcommand{\KL}{\mbox{KL}}  

\newcommand{\ie}{{\em i.e.~}} 
\newcommand{\vs}{{\em vs.~}}  
\newcommand{\eg}{{\em e.g.~}} 
\newcommand{\etal}{{\em et~al.~}}

\newcommand{\co}[1]{}

\newcommand{\cgrp}{\hat{g}}  
\newcommand{\rgrp}{g^o}  

\newcommand{\refset}{S^o}  
\newcommand{\cset}{\hat{S}}  

\newcommand{\match}{match}  

\newcommand{\gscore}{s^o}  

\newcommand{\Prec}{precision}  
\newcommand{\R}{recall}  

\newtheorem{prop}{Proposition} 

\title{Binomial Tails for Community Analysis}

\date{} 					

\author{\and Omid Madani \and Thanh Ngo \and Weifei Zeng \and Sai Ankith Averine 
  \and Sasidhar Evuru \and Varun Malhotra
  \and Shashidhar Gandham \and Navindra Yadav \\
  omadani|thanhngo|weifzeng|saverine|sevuru|varmalho|gshashi|nyadav@cisco.com \\
Cisco Tetration Analytics
}

\renewcommand{\headeright}{}
\renewcommand{\undertitle}{}

\hypersetup{
pdftitle={A template for the arxiv style},
pdfsubject={q-bio.NC, q-bio.QM},
pdfauthor={David S.~Hippocampus, Elias D.~Striatum},
pdfkeywords={First keyword, Second keyword, More},
}

\begin{document}
\maketitle

\begin{abstract}
An important task of community discovery in networks is assessing
significance of the results and robust ranking of the generated candidate groups.
Often in practice, numerous candidate communities are discovered, and
focusing the analyst's time on the most salient and promising findings
is crucial.  We develop simple efficient group scoring functions
derived from tail probabilities using binomial models.  Experiments on
synthetic and numerous real-world data provides evidence that binomial
scoring leads to a more robust ranking than other inexpensive scoring
functions, such as conductance. Furthermore, we obtain confidence
values ($p$-values) that can be used for filtering and labeling the
discovered groups. Our analyses shed light on various properties of
the approach.  The binomial tail is simple and versatile, and we describe two
other applications for community analysis: degree of community
membership (which in turn yields group-scoring functions), and the
discovery of significant edges in the community-induced graph.
\end{abstract}




\section{Introduction}
\label{sec:intro}

\co{
  . Read 10+ papers and see how they structure their intro, motivation, etc.!

  . general setting, our particular domain, motivation that started
  our work, description of contributions, org of the paper.
}

Automated discovery of candidate communities in networks finds a
variety of applications in physical and social sciences (biological and
biochemical networks, physical and virtual human networks)
\cite{fortunato2009,newman2010}. Given a graph representing binary
relations among nodes, informally and intuitively, a community
corresponds to a subgraph, \ie a subset of nodes, with relatively high
edge density among the community members (nodes of the subgraph), and
comparatively lower density of edges going outside the community.
Defining communities more precisely and what overall community
structure may be in various domains, and design of efficient robust
algorithms for uncovering such in networks has been the subject of much research
\cite{fortunato2009,fortunato2016}.

In our use-case, we are interested in the automated discovery and
effective presentation of candidate communities comprised of computers
(hosts) in an enterprise network.  In particular this effort is a
component of a tool that provides a user, such as a security
administrator of an organization, visibility into their complex
network, and importantly helps the user partition the network into
groups corresponding to geographic partitions, different departments,
and hosts running different applications in the organization.  This
partitioning and naming of the groups is a necessary step in defining
and maintaining network security policies, aka {\em network
  segmentation}: hosts in different groups (segments) can only
communicate on a few well-defined and restricted channels. Such policy
enforcement severely limits penetration and spread of malware and
hackers. This step of grouping hosts and assigning meaningful
names/labels to the groups, with the human in the loop, is also highly
useful in generating insights, for example in uncovering broad
patterns of communications with applications not just for security but
also for network optimization.

\co{
We posited that communities of hosts, as defined in the literature,
could be a good proxi for software applications and departments. We
have developed a tool that builds a graph from the network
communications and uses community discovery, and feedback from
preliminary use of the tool showed that the output, a list of groups,
is very valuable in helping the user partition their organization (set
of hosts).
}

Given the graph of communications among hosts, we find that while many
of the candidate groups generated, based on community detection,
correspond well to natural groupings (such as different departments,
employee groups, or applications), effective ranking and confidence
assignment to groups, along with other contextual information, is
crucial for the user to make sense of the output, as there are
numerous groups of varying quality and size generated.\footnote{Along
  with significant preprocessing of the graph, we use standard
  algorithms such as Louvain for community detection \cite{louvain}. }
In addition to better focusing the attention of the user, effective
ranking and labeling of the groups is critical for improved user
experience and bolstering user confidence in the tool.  In particular,
in our application, users are primarily interested in midsize and larger
groups, \ie preferably groups comprising 100s of members (hosts) and
beyond.

In the published literature, we find that there exist simple efficient
scores, such as conductance and triangle participation ratio
\cite{leskovec2012icdm}, but we observed that these were prone to rank
well-separated but rather small groups high. Other techniques such as
stability analysis can provide more robust results, but suffer from
several drawbacks: the problem of relatively small groups may still
surface, the methods can be expensive as they often require multiple
runs, and they depend on a choice of parameter settings and the
discovery algorithm and/or the graph perturbation specifics, which can
be hard to tune and explain to the user.



We explored a statistical approach based on binomial modeling, as a
principled approach to incorporating a 'natural bias' for larger
sizes, motivated by the intuition that larger groups tend to provide more
evidence and may therefore attain higher significance. Two of our key findings
can be summarized as: \begin{enumerate}
\item The tail of the binomial is a natural fit to the task.
\item There exists an adequate efficient approximation to the tail \cite{arratia89, ash90}.
  \end{enumerate}
The analyses, of the tail formula and the bounds, together with our
empirical findings show that original motivation is met well: binomial
scoring favors larger groups compared to other measures such as
conductance.
The score is equivalent to statistical significance ($p$-values),
which makes it convenient to (color) label the discovered groups (\eg
from 'Very High Significance' to 'Weak' and 'No Significance'), and to
filter groups when significance is too low.


Furthermore, we observe that the scores are not blunt (or too coarse):
ranking using the tail is quite competitive in various synthetic
settings, beating other inexpensive methods. It is also competitive on
a variety of real-world graphs we test on, including email graphs,
co-authorship, document similarity (for text clustering), and so on
(Section \ref{sec:exps}).  The binomial modeling space is rich, and we
explore a few variants and extensions, and discuss several properties
as well as potential limitations of the approach (such as coverage,
and significance \vs intensity of the community property).  Community
analysis is a complex multi criteria task, and we expect that binomial
modeling will be a useful complement to existing measures, such as the
simpler scores based on raw counts and simple ratios, and those based
on stability and graph perturbation.

We emphasize that binomial tail modeling is versatile and we
present two related applications in community analysis: 1) to assess
the strength of a node's membership, or node centrality with respect
to a discovered community \cite{newman2010,centComm2019}, we devise a
binomial score, which in turn yields additional methods for
group-scoring, and 2) we develop a binomial score for significance of
edges (interactions) among communities.  In all these cases, we spot a
pattern: a progression from using plain 'degree', such as a node or
group degree (a simple count), to a ratio (\eg conductance), to
binomial modeling (yields a probability and confidence).  Broadly, we
expect that the binomial tail is useful in other applications in
domains such as information retrieval.






%


This paper is organized as follows. In Section \ref{sec:form}, we
describe the binomial score, the efficient approximation to the tail,
and present an analysis of the growth of the score and the quality of
the approximation, and explore variants of the score. We also describe
and discuss the other scoring techniques we compare against,
including, in addition to conductance, the triangle participation
ratio, and the group-wise component of the Modularity objective (which
also favors group size) \cite{newman_2004,leskovec2012icdm}. Section
\ref{sec:exps} describes the evaluation methodology, and reports on
our extensive experiments, both in synthetic (planted partition)
settings as well as on real-world data.  Section \ref{sec:rel}
presents related work, and Section \ref{sec:summary}
concludes. Appendices contain further experiments on additional
synthetic settings and the different variants of the score, present a modularity
score based on number of nodes, and describe the two additional
applications of binomial tails for network analysis: evidence of
membership and significance of edges in the community-induced graph.

\co{
  In our application, we sought sizeable communities of 100s of nodes.

But several of the common existing methods normalize the value in a
way that size is ignored.

For instance, resistance is

larger groups were sought after for breaking into geographic areas,
departments, etc.

should we put this??  intuitively the larger the group, the more statistical
evidence as well.

We sought a simple principled scoring method for ranking that was
sensitive to size.

Furthermore, we find that this scoring provides a significance value
p-value.

Efficiency of evaluation

we develop two techniques..

We also highlight the application of the binomial modeling to two
other related tasks: scoring the 'degree' of membership, ie a measure
of strength of membership to a community, and scoring significant
edges among the discovered communities.

surveys: \cite{fortunato2016,fortunato2009}

Lets cite: \cite{leskovec2012icdm,leskovec2016snap,fortunato2016,leskovecNature2018}

}


\section{Formulation}
\label{sec:form}




An undirected unweighted graph $\G$ consists of a node set
$\nodeset$ and edge set $\edgeset$.  A group $\g$ is a subset of nodes
($\g \subseteq \nodeset$).  An edge (unordered pair of nodes) $\{u,v\}
\in \edgeset$ is referred to as an (incident) edge of the group $\g$
if $u \in \g$ or $v \in \g$.  We refer to the number of edges of a
group as the {\em degree} of the group (akin to degree of a node),
denoted by $\deg(\g)$. Thus, $\deg(\g) = |\{\{u,v\}\in \edgeset, \mbox{s.t. } \  u \in \g \ \mbox{or}  \ v
\in \g \}|$. To declutter notation, we often drop the generic group
$\g$ and simply write $\deg$, as we often do for other functions of
groups to be introduced.  If both ends of an edge are in the group
$\g$, we will call it an {\em internal} edge, and otherwise (exactly
one-end in $\g$) it's an {\em outgoing} edge. We refer to the number
of internal edges as the {\em internal degree}, and denote it with
$\din, \din \le \deg$.  Intuitively, a group that is dense inside and
sparse outside, \ie the observed ratio $\frac{\din}{\deg}$ is
relatively large, may correspond to a real-world community.




Now, given is a group (generated by some algorithm or process) with
degree $\deg$ and internal degree $\din$, and we are interested in the
probability of observing such an event by chance, which we formalize
next. We assume a uniform random model of connection as the {\em null}
(reference) model: for each of $\deg$ edges, one end is already in the
group, and the other end is picked from all vertices at random. Each
such pick is a random trial, so there are $\deg$ trials.  We define
the event of interest as: observing $\din$ or more internal edges, or
$\din$ of the trials pick a node in the group, out of $\deg$ trials.
If the probability of such event, according to the null model, is
tiny, or in other words it appears to be an extreme event, then we have
good evidence that the given group is special, in terms of density
inside relative to all group edges or outgoing edges (an indication of
community property), \ie this is unlikely to have happened by chance.
Note that, as is common when assessing statistical significance, we
are being conservative, and for example not asking for the probability
of the event of observing exactly (or around) $\din$ internal edges as
such point-events are often unlikely to begin with (there are many
possible values for $\din$).






Let $\pn$ be the (expected) probability that the other end of a group
edge is in the group (a 'success'), which according to the null model
is $\pn=\frac{\gsize}{\nodes}$ ('expected' according to the null
model). We call this model the {\em node-based} model (an edge-based
model is described in Section \ref{sec:edgeway}).\footnote{If we don't
  allow self-edges, we could use the more precise $\pn=\frac{|\g| -
    1}{|\nodeset| - 1}$.} We refer to $\frac{\din}{\deg}$ as the {\em
  observed proportion}.  The tail probability,
$\mathtt{BinomialTail}(\deg, \din, \pn)$, that we observe $\din$ or
more successes out of $\deg$ independent attempts, with success
probability $\p=\pn$, is given by:

  
\begin{equation}
\label{eq:binomial}    
  \mathtt{BinomialTail}(\deg, \din, \p) =  \sum_{i \ge d_{in}}
         {\deg \choose i} \p^i(1-\p)^{\deg-i}, \mbox{\ \ where, for the node-based model, }
         \p=\pn=\frac{|\g|}{|\nodeset|}
\end{equation}

The tail probability in a principled manner combines the connectivity
attributes of the group (degree and internal edges) and other context
(group and graph size). For weighted edges, if they are integers or
roundable to integers, the graph can be treated as a multigraph, and
the statistics such as degree appropriately computed by adding the
multiple edges (a similar approach to computing modularity on weighted
graphs).





\subsection{Evidence and Group Size}

Our original motivation was a principled method for scoring with a
preference for larger groups. Informally, larger groups should provide
more evidence of community structure assuming other properties such as
internal density remains roughly equal, since degree $\deg$ tends to
go up with larger groups.  However, the expected proportion $\p$
($\pn$) also goes up for larger groups, which can counteract $\deg$,
\ie the larger group could get a smaller significance.

For two isolated groups (no outgoing edges, or $\din=\deg$), where
nodes have similar (internal) degree in both groups, it is easy to
show that both situations can occur, as the following establishes. But
as long as the larger group is not too big, as a fraction of the
graph, indeed we obtain higher significance for the larger group:
\begin{prop}
  \label{prop:evidence}
For two groups with no outgoing edges, $\g_1$ and $\g_2$, $|\g_2| =
2|g_1|$, $\deg(\g_i)=\din(\g_i)=|\g_i|$ (node degree of roughly 1), the larger
group $\g_2$ obtains a lower tail probability (higher significance) if
and only if $|\g_1| < 1/4|\nodeset|$.  More generally, with
$|\g_2|=k|\g_1|, k > 1$, and $\deg(\g_i)=a|\g_i|, a > 0$, $\g_2$
obtains a higher significance iff $\frac{|g_1|}{|\nodeset|} <
k^\frac{-k}{k-1}$ (independent of $a$).
\end{prop}

For the proof, first for $k=2$, let $n=|\nodeset|$ and $|\g_1| = pn$,
thus $|\g_2| = 2pn,$ and $\deg(\g_1)=\din(\g_1)=|\g_1|=pn$, and
$\deg(\g_2)=\din(\g_2)=2pn$, and, from equation \ref{eq:binomial}, we
want to see when $p^{pn} > (2p)^{2pn}$, or $pn\log(p) > 2pn\log(2p)
\Rightarrow \log(p) > 2\log(2p)$, or $-\log(p) > 2\log(2) \Rightarrow
p < 1/4$. For general $k > 1$, replace $2$ with $k$ and insert $a$:
$p^{apn} > (kp)^{akpn} \Rightarrow \log(p) > k\log(kp) \Rightarrow \cdots$.

Similar patterns and proof applies to dense (isolated) groups such as
(near) cliques, \eg replace number of group edges with $0.5(pn)^2$,
and in this case the condition on the size of the larger group to
obtain higher significance becomes less strict (plausible, as $\din$
goes up quadratically).  Appendix \ref{app:resolution}, shows another
example, to illustrate a resolution limit problem for the binomial
tail, where a larger group scores higher although it may be almost
half as dense (but with more total number of internal edges). In
practice, it's also the case that all or most groups discovered are a
small fraction of the graph in number of nodes, and we empirically
find that the tail is typically well correlated with graph size in the
experiments of Section \ref{sec:exps}. The approximate bounds below
also show that generally as $\deg$ increases, the score increases as
well.


\co{
We note that for two groups $\g_1$ and $\g_2$ with the same expected
and observed proportions (\ie $\p_1=\p_2, \q_1=\q_2$), the one with more
edges (group degree), also gets a higher significance for most ranges
of $d$. For example, both group may have 10 nodes, in a 100 node
graph, thus $\p_i=0.1$, and $\g_1$ can have $10$ inside and $10$
outgoing ($\q_1=0.5$), while $\g_2$ has 20 internal edges and 20
outgoing ($\q_2=0.5$ as well). This is seen from the upper and lower
bound expressions of \ref{eq:kl}, presented next.
}

\subsection{Approximation}

Equation \ref{eq:binomial} is simple to compute, but it is inefficient
for many generated groups in large graphs: $O(k|\edgeset|$ for $k$
groups, assuming Stirling's approximation is used for factorials in
${\deg \choose i}$, and it can lead to numeric (floating point) problems.
Indeed in our experiments when we use the exact formulation we see
considerable slow down, of two orders of magnitude, even on relatively
small graphs of 100s of nodes (Appendix \ref{app:syns}).  Efficient
computation also aids other applications of the tail such as node
membership (Appendix \ref{app:apps}), and a potential use for repeated
evaluation as an objective for community mining. The approximate
bounds below also shed light on the properties of the tail such as on
group ranking and growth, as we will see.

There exists a normal approximation to the binomial tail
\cite{experimenters}, and one could use generic bounds such as
Chebychev \cite{MotwaniR95}, but these proved to be inadequate for our
use case (based on experiments). The normal approximation works best
for expected proportions $\p$ near $0.5$ and when the number of trials
is very large \cite{experimenters}. On real world datasets, we find
that the expected proportion $\p$ is often relatively small (\eg below
0.10) and the observed (connection) proportion $\q=\frac{\din}{\deg}$
is relatively large.  Fortunately, the following upper and lower bound
on the tail probability is sufficiently tight (see Section \ref{sec:good})
and serves us well
\cite{arratia89, ash90}:










\begin{align}
\nonumber  & \frac{1}{ \sqrt{2\deg} } \U \le \mathtt{BinomialTail}(\deg, \din, \p) \le \U, \mbox{ where } \\
  \label{eq:kl}
& \U = \exp(-\deg \KL(\q || \p)), \mbox{ where } \q=\frac{\din}{\deg} \mbox { \ \ \  (U is the upper bound on the tail probability), }
\end{align}


where $\KL()$ is the (asymmetric) relative entropy function (or
Kullback–Leibler divergence): $\KL(q|| p) = q\ln \frac{q}{p} +
(1-q)\ln\frac{1-q}{1-p}$, where $\ln()$ denotes the natural logarithm
function.  The ratio of the observed to expected proportion, $\i(\g)
=\frac{\q(\g)}{\p(\g)}$, is handy in the analyses below, and we will
refer to it as the {\em intensity} of the group (intensity of the
community property).


Thus we simply need the readily available number of graph nodes, size
of the group, and its degree and internal degree to compute the
approximation to the tail, in O(1) time.

\subsection{Binomial Scores}


For convenience of scoring and comparisons, as the probabilities can
get tiny for good groups, we will use the negative log base 10 of the
above upper and lower tail probabilities, akin to Richter scale for
measuring the power of earthquakes (see \ref{sec:range} below), and
take the average for scoring and ranking the groups:

  \begin{align}
\label{eq:approx} & \lscore(\g) =  \deg\KL(\frac{\din}{\deg}|| \p)/\ln(10) \hspace*{1in} \mbox{lower score of group $\g$, using the upper bound on tail probability} \\
\nonumber & \uscore(\g) =  \lscore(\g) + 0.5 \log_{10} (2\deg(\g)) \hspace*{0.8in} \mbox{upper score of group $\g$, via the lower bound on tail probability}
\end{align}


In our implementation, when $\deg \le 50$ we use the exact formula of
equation \ref{eq:binomial}, and otherwise the approximations from above:
\begin{align}
\label{eq:bn}
  \bsn(\g) = \begin{cases}
  -\log_{10}( \mathtt{BinomialTail}(\deg(\g),\ \din(\g),\ \pn(\g))),  \mbox{ \ \ \  when $\deg(\g) < 50$ } \\
  0.5(\uscore(\g) + \lscore(\g)), \hspace*{0.5in}  \mbox{ otherwise, where $\p = \pn = \frac{|\g|}{|\nodeset|}$
    \ \ \ {\em node-based binomial score} } 
  \end{cases}
\end{align}
Furthermore, if the observed ratio $q$ is not higher than the
expected, \ie when $q=\frac{\din}{\deg}\le \p$, we return 0 score
(insignificant).

Note that the arithmetic mean of upper and lower score corresponds to
(the negative log of) the geometric mean of the bounds on
probabilities. Next we use the upper and lower bounds to analyze the
accuracy of the approximation.



\subsection{Range and Growth of Binomial Scores}
\label{sec:range}
Looking at the upper and lower bounds of equation \ref{eq:approx}, we
see that the score can grow roughly linearly with group degree $\deg$,
and in the experiments we do observe such ranges for the higher
scoring groups. Moreover, we also observe a positive correlation with
group size as anticipated (larger groups tend to have higher
$\deg$). The score also goes up with KL divergence and in particular
increases in proportion with log of the intensity (so at a slower rate
compared to $\deg$), as the following proposition formalizes.

\begin{prop}
\label{prop:incr}
  With $0 < \p < \q \le 1$, KL divergence increases with increasing
  intensity $\i=\q/\p$. Therefore the approximate binomial score
  (either the upper or lower or their mean) increases with either
  increasing $\deg$, when KL remains the same or also increases, or
  with increasing (log) intensity $\i$, when $\deg$ remains the same.
\end{prop}

The first term of KL divergence, $q\ln \frac{\q}{\p}=q\ln \i$, is
increasing in intensity $\i$, but the 2nd term ($(1-\q)\ln\cdots$) is
(slightly) decreasing.  To show that KL increases with $\i$, we will
take the derivative and show the derivative is positive, under two
cases, that either $\p$ or $\q$ is constant.  For fixed $\q$,
replacing $\p$ by $\q/\i$, $\frac{\partial \KL(\q||\p)}{\partial
  \i}=\frac{\partial}{\partial \i}(q\ln(\i)) +
\frac{\partial}{\partial \i}[(1-\q)\ln(\frac{1-\q}{1-\q/\i}] =
\frac{\q}{\i} + (1-\q)\frac{1-\q/\i}{(1-\q)}(\frac{-\q}{\i^2}) =
\frac{\q}{I}(\frac{\i - 1 + \q/\i}{\i}) > 0$ ( when $0<\p<\q\le1$,
thus $\i > 1$). Similarly, for fixed $\p$ (varying $\q < 1$),
$\frac{\partial \KL(\q||\p)}{\partial \i}=\frac{\partial}{\partial
  \i}[\p\i\ln\i + (1-\p\i)\ln\frac{1-\p\i}{1-\p}]= (\p\ln\i + \p) +
[-\p\ln\frac{1-\p\i}{1-\p} +
  (1-\p\i)(\frac{1-\p}{1-\p\i})(\frac{-\p}{1-\p})]=\p(\ln\i-\ln\frac{1-\p\i}{1-p})
> 0$, as the remaining second term
$-\ln\frac{1-\p\i}{1-p}=-\ln\frac{1-q}{1-p} > 0$, with $\p < \q$.

That the binomial score should increase with $\q/\p$ (fixed $\deg$,
increasing $\din$) is perhaps easier to see with the exact formula of
Equation \ref{eq:binomial}, as the set of terms to sum over (all positive) strictly
shrinks as $\din$ is raised, but it's proportional dependence on
$\deg$ is easier to see here.  As a concrete example of equal
intensity but different degrees, two groups $\g_1$ and $\g_2$ each
having 10 nodes, in a 100 node graph, thus same $\p_i=0.1$, $\g_1$
having $10$ inside and $10$ outgoing edges ($\q_1=0.5$), with $\g_2$
having 20 internal and 20 outgoing edges ($\q_2=0.5$ as well), $\g_2$
obtains almost twice the score.

Since we use log base 10, the scores 1, 2, 3, $\cdots$ correspond
respectively to chance (tail) probabilities of 0.1, 0.01, 0.001,
$\cdots$ thus a score of 1 corresponds to a fairly weak confidence
(0.1 p-value), and each successive increment implies 10 times more
significance (akin to the Richter scale), and scores around 2 and
higher indicate increasingly strong confidence. In practice, we
observe many generated groups with scores well above 3, and basically
the score growing proportional to group degree $\deg$ for many of the
discovered groups.


\subsection{Goodness of the Approximation}
\label{sec:good}

Scoring and ranking need not be so precise as we only use approximate
models of reality to assess significance, and often community
discovery tasks have an inherent subjective or imprecise component.
However, an approximation that is too far from the intended model
could defeat the initial motivation for using that approach.  Here we
will see that the approximation quality is promising, and furthermore
as the community property goes up (more precisely, when the
approximate lower and upper scores increase), the error in
the approximation also rapidly shrinks.

A good way to assess the quality of the approximations above is to
look at the relative error, which can be achieved since we have both
an upper and a lower bound: Let $s(\g)$ be the actual binomial score
of group $\g$, and denote the relative error by $r(\g), r(\g) =
\frac{|s(\g) - \bsn(\g)|}{s(\g)}$ (for $s(\g) > 0$).  We can readily
see the following ($10\%$ or better) guarantee on the quality of the
approximation holds:

\begin{prop} Let $\tilde{r}(\g) = \frac{\uscore(\g) - \lscore(\g)}{\lscore(\g)} = \frac{0.5
    \ln(2\deg)}{\deg\KL(\q|| \p)}$. We have $r \le \tilde{r}$ and 1)
  $\tilde{r} < 0.1$ when $\KL() \ge 0.5$ and $\deg \ge 50$, and 2)
  (more generally), with $0 < \p < \q \le 1$, error $\tilde{r}()$ is
  reduced with increasing degree $\deg$, with intensity fixed or
  increasing, or increasing intensity $\i$ (with degree fixed or
  increasing).
\end{prop}

The first claim, $r(\g)\le \tilde{r}(\g)$, follows from bounding the
numerator of $r(\g)$, \ie noting that $|s(\g) - \bsn(\g)| \le
\uscore(\g) - \lscore(\g)|$ (as both $s(\g)$ and $\bsn(\g)$ are in
$[\lscore(\g), \uscore(\g)]$), and for the denominator, $\lscore(\g)
\le s(\g)$. The second part follows from the expression for
$\tilde{r}$ and that KL increases with intensity from Proposition
\ref{prop:incr}.

\co{ When we look at a worst case version of relative error, \eg the
  ratio of the difference of the upper and lower score to the lower
  score, $\frac{\uscore - \lscore}{\lscore} = \frac{0.5
    \ln(2\deg)}{\deg\KL(\frac{\din}{\deg}|| \p)} $, we see that, as
  long as KL divergence is 0.5 or higher, the worst-case error,
  rapidly goes to 0 as degree grows, and, for example, for degree
  $\deg \ge 50$, the relative error is already below 10\%.  }

We use the exact expression for the score when $\deg \le 50$, thus we
framed the above claim for $\deg \ge 50$. The higher the degree the
lower the relative error, but we will focus briefly on the KL
divergence component of the error to see it is above say 0.5. The
divergence is above 0.5 (and easily exceeds 1) when the intensity $\i
\ge 2$, {\em and} $\p$ is relatively high, \ie $\p \ge 0.1$, \eg when
$\p = 0.1$, we need $\q \ge 0.21$, for a guaranteed below 10\%
relative error, when we ignore the additional help from $\deg$, and
error goes to below 5\% for $\deg\ge 100$. However, for a constant
intensity, the KL divergence goes down as $\p \rightarrow 0$.  For
instance, for $\p=0.01$ (a group of size hundredth of the graph), KL
divergence is above 0.5 only when $\q \ge 0.056$ (a requirement that
$\i \ge 5$).  In practice (on various datasets) many of the discovered
candidate communities do have high intensities, and one can argue that
more accurately scoring small groups with relatively low intensities
(low quality groups) is not as critical (see also Appendix
\ref{app:syns}).

For smaller groups, degree $\deg$ tends to also be small,\footnote{In
  particular, in (unweighted) sparse graphs (or
  $|\edgeset|=O(|\nodeset|)$), $\deg$ is a multiple of group size
  $\gsize$, assuming each node has roughly the average degree. } and,
looking at both the upper and lower bound on the score, we require a
higher intensity to attain a comparable score. This observation is
consistent with the intuition that the smaller the trials (the
evidence), the smaller the community quality score of a group should
be (see also Proposition \ref{prop:evidence}).

A tighter
(smaller) bound on relative error, as we use a midpoint, is:
$\max(\frac{\bsn(\g) -
  \lscore(\g)}{\lscore(\g)},\frac{\uscore(\g)-\bsn(\g)}{\bsn(\g)}))$,
but the above simpler expression gives an idea of the quality of the
approximation.  In practice, we could also use only the conservative
lower score, and also report the actual worst-case error $\tilde{r}()$
when we use the approximate binomial score. Our ranking experiments
showed little difference between using the lower or the average score,
as well as, in most cases, no difference in ranking quality when using the
exact \vs the approximate form of the score  (Appendix \ref{app:syns}) and
\ref{app:bin}).



\subsection{A Null Model Based on Edge Counts}
\label{sec:edgeway}


In this null model, like above we are given a group with some number
of incident edges $\deg$, where we are free to pick the other end of
each such edge. Like above, we pick the other end of each such edge
independently with some probability $\p$, and $\p$ is different for
different groups. Here, the probability $\p$ is a function of the
number edges of the group $|\deg|$ and of the graph $|\edgeset|$ (\vs
the number of nodes of the group $|\g|$ and graph $|\nodeset|$), or
the relative 'chattiness' of the group, so a rough measure is
$\frac{\deg}{\m}$. More precisely, we use the following small variant,
which we denote by $\pe$:
\begin{equation}
\pe = \frac{\deg + \din}{2\m}
\end{equation}

To motivate the small difference in the definition of proportion,
imagine for each edge of the graph, we look at one end, randomly
picked, to see whether or not it's in group \g. Then the expected
number of nodes in $\g$ that we observe is $\frac{\deg + \din}{2}$
(out of $\m$ observations).  This is the same null model behind the
modularity objective \cite{newman_2004}, where the expected proportion
of internal edges, \ie both ends in $\g$, comes out to
$\pe^2$.

We can replace $\pn$ with $\pe$ in Definition \ref{eq:bn} to get the
edge-based binomial score $\bse(\g)$, \ie the main difference is the
choice of the observed proportion $\p$. Note that $\frac{\deg +
  \din}{2}$ could also replace the number of trials $\deg$ in
Definition \ref{eq:bn}.

Symmetrically, one could define a modularity objective based on number
of nodes, and a corresponding Louvain algorithm based on node
proportion \vs edge proportion. See Appendix \ref{app:node_louvain}.

\subsection{A Comparison of the Edge and Node Based Models}
\label{sec:two_models}

\begin{figure}[!htbp]
\begin{center}
\leavevmode   
\subfloat[Graph with two groups ($\nodes=8$, $|\edgeset|=9$)]
         {\includegraphics[width=5cm,height=2.5cm]{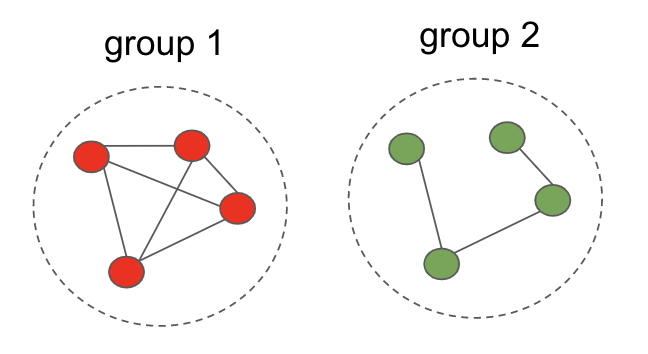}}
\subfloat[More nodes (without) added to the graph]
         {\includegraphics[width=5cm,height=2.5cm]{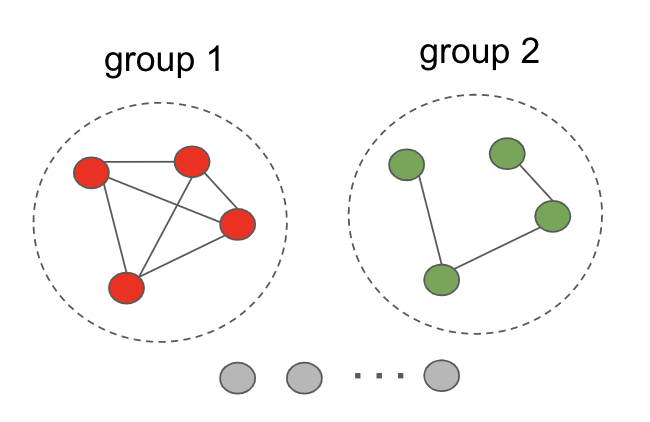}}
\subfloat[Outgoing edges added to group 1]
         {\includegraphics[width=5cm,height=2.5cm]{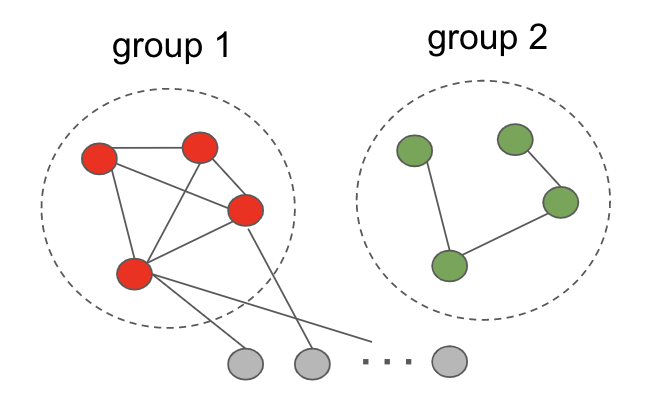}}
\end{center}
\caption{Example graphs where the node-based binomial model 'prefers'
  group $\g_1$ (\ie scores group 1 higher), but the edge-based model
  prefers group $\g_2$. In all three cases, $\pn(\g_1) = \pn(\g_2)$,
  and for graphs in (a) and (b), for both groups and in either node-
  or edge-based model, the tail probability is $\p^{\deg(\g)}$ (as
  $\din = \deg$).  See text for further explanations.}
\label{fig:comparare_null_models} 
\end{figure}

What is the consequence of using one null model \vs the other?  The
implications are easier seen when we look at groups that have no
outgoing edges, \ie $\deg=\din$: the edge-based model may prefer
(internally) less dense groups (as $\pe$ can be smaller for
less dense group, irrespective of group size), compared to the
node-based model, which may prefer smaller (but possibly denser)
groups (as $\pn$ will be smaller or the exponent can be higher).
To illustrate this difference,
we present an extreme example in Figure
\ref{fig:comparare_null_models}(a).  For both groups, in either null
model, following equation \ref{eq:binomial}, the tail probability is
$\p^{\deg(\g)}$ (as $\din = \deg$). For the node-based model,
$\p_n(\g_1) = \p_n(\g_2) \implies \p^6_n(\g_1) < \p^3_n(\g_2)$ (where
$\pn(\g_1) = \frac{|\g_1|}{|\nodeset|} = \frac{4}{8} = \pn(\g_2)$).
For the edge-based model, $\p_e(\g_1)=\frac{2}{3} >
\p_e(\g_2)=\frac{1}{3}$, and $(\frac{2}{3})^6=(\frac{4}{9})^3 >
(\frac{1}{3})^3$. In this example, the node-based model prefers the
group that looks more like a community (both groups are isolated, but
group 1 is denser).

The node set $\nodeset$ grows in Figure
\ref{fig:comparare_null_models}(b), but no new edges are added, and
since $\p_n(\g_i)$ shrink ( have $\p_n(\g_1) = \p_n(\g_2)$), we see
that the significance of both groups under the node-based null model
(binomial score) grows, but score grows faster for group 1, while
there is no difference in the groups' scores in the edge-based
model.  Finally, in Figure \ref{fig:comparare_null_models}(c), as
outgoing edges are added to $\g_1$, the preference of $\g_1$ over
$g_2$ decreases, though $\g_1$ may still be preferred: this depends on
the 'value' of an internal edge \vs an outgoing edge, which in turn
depends on $|\nodeset|$ relative to $|\g_1|$ (or $\pn$). On the other
hand, in the edge-based model, the preference of $\g_2$ over $\g_1$
only grows. We can see that the edge-based model can have an advantage
now: it can prefer relatively isolated groups with good density, while the
node-based model may prefer smaller dense groups that nevertheless
have many outgoing edges (less community-like to the naked eye).


We have observed on several datasets that, as might be expected, the
rankings by the two techniques are highly correlated (\eg Spearman
rank correlation of over 0.99 on 5000 groups from Amazon co-purchase
dataset in SNAP \cite{leskovec2016snap}).  In another experiment, on a
few graphs derived from network communications, we observed that if we
take the maximum of the two scores, and use 1.5 as a threshold on
score (corresponding to roughly 95\% confidence), we observed that we
get 5\% more groups deemed significant than if we use either measure
in isolation.

%

\subsection{A Global Bound}
\label{sec:global}

The above null models were conditional on an observed group, \ie {\em given} a group with
certain size $k$ and degree degree $\deg$ (event $A$), we sought the
probability of observing $\din$ or more internal edges (event $B$).
The graph can be big and it may not be so surprising (so unlikely) to
observe such a group in a large graph. Here, we want to add more
context and account for graph size. The union bound (Boole's
inequality), \ie the probability of a disjunction is bounded by the
sum of component probabilities \cite{MotwaniR95}, gives us an upper bound on the
probability of the event $C$, that we observe {\em one or more groups
  (\ie subsets) of size $k$ with degree $\deg$ and internal degree $\din$ or more}:
\begin{align}
\nonumber P(C) & \le \sum_{\g \subseteq \nodeset, |\g|=k}  P(A) P(B|A), \mbox{\ from the union bound } \\
\nonumber P(C) & \le {|\nodeset| \choose k}  P(B|A) \mbox{\ \ \ \ \ \ (as $P(A) \le 1$) } \\
\label{eq:gb}  & = {|\nodeset| \choose k} \mathtt{BinomialTail}(\deg, \din, \p)
\end{align}
where $P(B|A)$ is given by equation \ref{eq:binomial}. We can define a
corresponding score, {\em global} binomial score, by taking the
negative log. We note that $\ln({|\nodeset| \choose k})$ can be
efficiently approximated using Sterling's formula for the factorial.

In our experiments, on synthetic datasets with relatively small groups
and graphs, the global binomial score goes to $\le 0$ (as the upper
bound of \ref{eq:gb} reaches and exceeds 1.0), while the binomial
effectively ranks the groups. On the real-world graphs data sets, we
did not observe a significant difference between the two (see Appendix
\ref{app:bin}).\footnote{To get a somewhat tighter bound, instead of
  replacing with 1.0, one could replace $P(A)$ with probability of the
  event of observing a size $k$ group with degree $\le \deg$. This
  requires a further generative assumption on the overall graph and
  does not purely reward community structure anymore, but how unlikely
  a group size and degree is with respect to the choice of graph
  generation. We didn't observe a difference when we used the uniform
  Erdos-Reny model for graph generation estimating $P(A)$.  } Ixn
practice, one can use the more conservative bound of \ref{eq:gb} for
filtering and/or labeling and ranking candidate groups, with a
possible fall-back to plain binomial score for ranking the borderline
cases.

\subsection{Other Scoring Methods}
\label{sec:methods}

We compare ranking via the binomial score $\bsn(\g)$ with several simple
scoring methods:
\begin{enumerate}
\item Modularity (group-wise) \cite{newman_2004,leskovec2010}: $\qi(\g) = \frac{\din(\g)}{\m} - p_e(\g)^2$, is the
  fraction of observed internal edges minus a expectation for such
  fraction, according to the null model of Section \ref{sec:edgeway}.
\item Conductance \cite{malik2000}: $\frac{ \deg(\g) - \din(\g) }{ \deg(\g) + \din(\g) }$ (the lower the better).\footnote{In the denominator, the minimum of degree and internal degree for $\g$ and its complement
  (rest of the graph, $\nodeset - \g$) is used.
  Since $\g$ is substantially smaller than half of graph in our experiments,
  we do not see a difference in conductance value.  }
\item Triangle participation ratio (TPR) \cite{leskovec2012icdm}: for a group $\g$:  $TPR(\g) =
  \frac{|\{u: \exists t \in triangles(\grp), \ u \in t \}|}{|\grp|}$,  where \\
  $triangles(\grp)= \{\{u, v, w\}:\{u, v, w\} \subseteq \grp, \{\{u,v\}, \{v,w\}, \{w,u\}\} \subseteq \edgeset \}$
\item Descending group size (number of nodes in the group, Size)
\end{enumerate}




We report correlation with group size per our motivation to bias
rankings towards larger groups. We also compare with ranking using
plain group size (largest first) as a baseline because in several
datasets and parameter settings this baseline (or its reverse) does
very well and sheds light on the nature of the specific ranking
problem.  We also report on the reverse (smallest first) in a few
cases.

We compare against $\qi()$ since we expect $\qi()$ to correlate with group
size as well, and it shares some similarity with the Binomial score:
it is also a function (the difference) of a candidate group's internal
edges \vs an expectation of such.  It can be readily available when we
use the Louvain algorithm and has been used in ranking comparisons
\cite{leskovec2012icdm,leskovecNature2018}. We note that the original
modularity formulation sums per-group differences over all groups to
assess an entire partitioning while we are using the group-wise
component of modularity for ranking groups.

Among a variety of simple scoring techniques tested, Yang and Leskovec
found that conductance and TPR performed among the best (one
reflecting internal density, another reflecting external separation)
\cite{leskovec2012icdm} and conductance in particular has been used
often as an objective for community discovery
\cite{seedSets06,fortunato2009}. Note that among these methods, only
TPR requires some nontrivial algorithm to compute.

\subsubsection{Discussion}
\label{sec:discuss}

\begin{figure}[!htbp]
\begin{center}
\leavevmode   
\subfloat{\includegraphics[width=15cm,height=2.5cm]{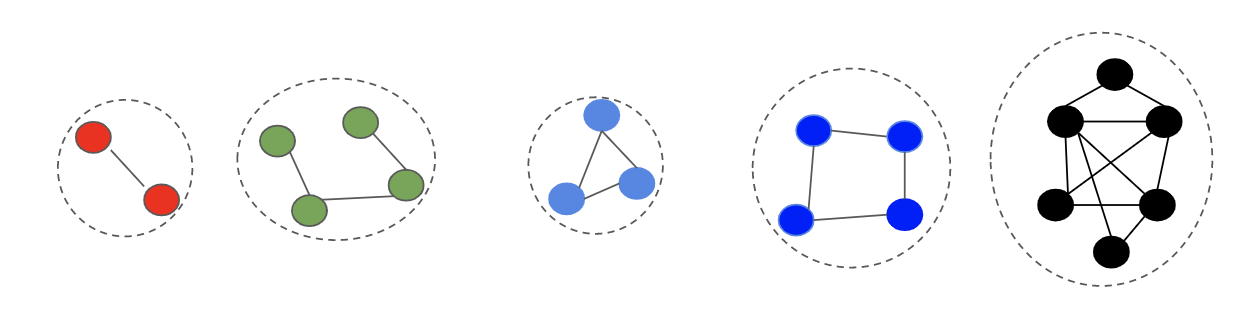}}
\end{center}
\caption{Examples where conductance and TPR show no discrimination,
  eventhough a larger or a denser group is often  preferred.  All groups are isolated (no
  outgoing edges) and have perfect conductance of 0. The first two on
  the left have 0 TPR (no triangles), and the remaining three groups
  have a perfect TPR of 1.0.}
\label{fig:examples} 
\end{figure}

Each technique has pluses but also shortcomings. TPR is an imperfect
measure of density inside the group and ignores outgoing edges. Note
also that TPR can be a perfect 1.0, and yet the internal density
($\din$) of a group can have much room for improvement. We see the
issue of tied scores often for TPR in the experiments. Both TPR and
conductance are 'normed' measures that factor out (ignore) group size
or number of connections (statistical evidence). Figure
\ref{fig:examples} show several groups with the same (perfect)
conductance of 0.0 and/or perfect TPR of 1.0 with varying sizes and
internal densities. Thus both methods can be insensitive to size and
internal density in some circumstances.  Under stability and
perturbation analysis tools, these structures may also obtain similar
scores since they may be rediscovered under multiple runs.  But
intuitively or at least in some tasks (like ours), one prefers larger
sizes and higher internal densities. We saw that the binomial score is
sensitive to increasing internal edges as well as overall degree.



The binomial score provides a principled way of combining internal and
outgoing edges and taking into account some of the other contextual
information (group and graph size). However, the approximations
inherent in using the tail, in equations of \ref{eq:approx}, or in the
union bound of equation \ref{eq:gb} may weaken the testing
(discriminative power and coverage). We observe these possibilities in
certain extreme cases in synthetic experiments (Appendix
\ref{app:syns} and \ref{app:bin}). It is also possible that real-world
groups enjoy high evidence, easily passing the significance threshold,
and for those passing the significance threshold (\eg a score above 2
or 3), the binomial score may become a blunt instrument for ranking
purposes.  We conduct experiments on synthesized and real-world
networks to shed light on these possibilities and some of the issues
will be explicitly raised and discussed. Note that intensity score of
$\i(\g) = \frac{\q(\g)}{\p(\g)}$ is another measure of strength of the
(community) property. This score ignores the raw degree counts (the
support or the evidence), akin to conductance and TPR. As an example,
a group A may have a high intensity of 10 ($\frac{q}{\p}=10$), but a
binomial score of only 2, while another group B may have an intensity
of 2, but high binomial score of 5: group B is larger than A, and has
many internal edges (providing much statistical evidence of community)
but also external edges, and subjectively, to the human eye, group A
may seem more of a community.  However, using intensity alone can
become too biased toward small or insignificant groups (this measure
is called ratio modularity in \cite{leskovec2010}).  In practice,
likely multiple measures are needed and should be presented to the
user. For example, one possibility is to filter by a confidence
threshold, and rank what remains by a measure of intensity.  We leave
exploration of intensity and combinations to future work. See also
\cite{leskovecNature2018} for combination techniques.

All these measures ignore how the internal connectivity is spread
among the group members. For instance, a given group might have a
small internal cut, \ie may be separable into two good nearly-sized
groups via removing only a few edges (a dumbbell shape, imbalanced
group).  Or some single or few nodes may have most of the internal
connections of a group (\eg star or core-priphery structure), while
intuitively a more balanced spread of the internal edges is
preferred. The scoring techniques do rely on the group generation
technique to discover a few good quality groups, possibly mixed with
lesser quality ones, and the task is to highly rank the best from among
what is given, where what constitutes 'best' can be application
specific.  See also Appendix \ref{app:resolution} on scoring groups
based on aggregating the strength of membership-level scores of
individual nodes, which, as explained there, may lead to more balanced
groups.

\co{
  
To discuss, some of the issues:
\begin{itemize}
  \item the problem with normalization: ignores size for our motivation
  \item Tied scores (several scoring techniques can exhibit this
    problem in different situations)
  \item Binomial score, range of valid values (power?),
    evidence/significance vs 'extremeness', could be too conservative
    (if we use lower bound), etc..? Also, it's for the overall group
    (some vertices may not really belong, or the could be small
    cuts), we are at mercy of the group generation technique here...
  \item Is there a problem to solve: maybe they all do well in
    practice?? Also, lack of groundtruth issues??
\end{itemize}
}

\section{Experiments}
\label{sec:exps}


We conduct experiments on both synthetic and several diverse
real-world datasets shown in Table \ref{tab:datasets} (social
networks, co-purchases, transportation, text clustering, $\cdots$).
On synthetic data, we follow the planted partition model and 'plant'
the groups.

The real-world graphs come with established groupings, often according
to some node attribute(s) or other external knowledge such as explicit
groupings by users. We refer to these groups as {\em reference} groups
(as opposed to groundtruth).  Researchers have questioned the utility
of treating such groups as groundtruth in assessing community
discovery algorithms \cite{gt2017}. We agree that groups discovered
via community analysis could be valid and highly useful and at the
same time be far from the provided groups.  Nevertheless, we expect
that better correlation with these reference groups, on {\em diverse
datasets}, can signal the robustness of a scoring method.  We also
critically examine and discuss the findings in some detail (\eg
reference groups can be coarse and general, while the discovered
groups can be relatively fine-grained, or vice versa).  For each
dataset, we briefly describe the nature of the graph and the reference
groups, and the reader can assess for herself the extent to which
interactions that determine the graph may correlate with the provided
reference groups.

In this section, we use the node-based binomial score, equation
\ref{eq:approx}, from the binomial family. Appendix \ref{app:bin}
reports on experiments with variants, including the edge-based
approach, and a $p$-value expression derived by He \etal
\cite{he2018detecting}.  We used the Louvain algorithm for group
generation \cite{louvain}, which we found to be fast and robust. We
picked the discovered groups from the final pass, where modularity is
highest. We also tried implementations of the Spectral algorithm, one
from sklearn.cluster \cite{scikit-learn}, but the algorithm did not
converge in several scenarios, such as when noise was increased in the
synthetic experiments (or the groups generated were far inferior). We
report ranking comparisons only on groups of minimum size 3,
consistent with \cite{leskovecNature2018}.

\co{

  'ground-truth'

We will call reference groups.

Discussion of why 'reference' vs 'groundtruth' ...

We hope that even though the reference groups are 'noisy' (better way
of saying it) there is good correlation ... 

Louvain vs 

}

\subsection{Evaluating Scoring Techniques}

A community discovery algorithm generates a set $\cset$ of {\em
  candidate} groups. We also have a set of reference groups $\refset$
(the planted groups in synthetic experiments or the provided reference
groups in the real-world datasets).  Note that we are concerned with
ranking the groups in the given $\cset$ set (and not on generating
better such sets).

We assess the performance of a candidate group output by a community
discovery algorithm by how well it overlaps the best matching group in
the reference set of groups. In particular, we define the following
{\em overlap score} for each candidate group:

\begin{align}
  \label{eq:quality}  \gscore(\cgrp) = \max_{\rgrp \in \refset} \match(\cgrp, \rgrp ),
  \mbox{\ \ \ where }
  \match(\cgrp, \rgrp )  = \sqrt{\R(\cgrp, \rgrp) \ \Prec(\cgrp, \rgrp )}, \\
\hspace{-2in} \mbox{ where } 
\nonumber  \R(\cgrp, \rgrp)  = \frac{|\cgrp \cap \rgrp|}{|\rgrp|}, 
  \ \Prec(\cgrp, \rgrp)=\frac{|\cgrp \cap \rgrp|}{|\cgrp|} 
\end{align}

Thus we use the geometric mean of 'precision' and 'recall' as the
score (the ratio of intersection size to candidate group size is
interpreted as the precision score, common in supervised learning),
and the score of a candidate group is the highest such score among the
match attempts against all the reference groups. Geometric mean is
more conservative than the arithmetic mean (it is closer to the lower
of the precision and recall scores). We could also use the simpler
Jaccard index \cite{jaccard1901} (\ie $\frac{|\cgrp \cap
  \rgrp|}{|\cgrp \cup \rgrp|}$), but we had a small preference for the
mean since Jaccard can be lower than both the precision and recall
scores, while the mean indicates that at least one score is not lower.
Note also that the matching is not exclusive: the same reference group
can match multiple candidate groups.\footnote{Note that number of
  groups in reference and candidate lists can differ widely.  To
  simplify and for efficiency of evaluation (so that matching order
  wouldn't matter), we don't constrain that the matching be one to
  one, even if it were possible. Thus a reference group, can yield
  best match score to multiple candidate groups.  Since the candidate
  groups are mutually exclusive, at most one candidate group can get a
  relatively high overlap score out of the several that match the same
  reference group. }

When we sort the candidate groups based on descending order of overlap
score $\gscore()$, we obtain the {\em reference ranking}.  For each
scoring method (binomial score, conductance, \modularityr, $\cdots$),
we rank the candidate groups by the scoring method, and report the
Spearman's rank-order ({\bf SPR}) correlation \cite{spearman1904}, of
this candidate ranking with the reference ranking (specifically, between
the two score lists), similar to the experiments of Zitnik \etal
\cite{leskovecNature2018}. An example of evaluating is given in
Section \ref{sec:example}.  We also report the overlap score
$\gscore()$ of the top ranked group, averaged over the trials (in case
of ties, at the top, we use the average of those tied), denoted {\bf top PR}.
For larger datasets, when there are many candidate groups generated,
we also report the average $\gscore()$ of top 5 for each method, {\bf
  top 5 PR}. SPR is useful to obtain an impression of the entire
ranking, while in some settings we are only interested in how the top
one or few groups are performing.  We also report the average overlap
score (average 'performance'), denoted \avgPR, of the generated
candidates groups, so that the overlap score of the top ranked is put
in context: We expect that, for a good scoring technique, the top
ranked should have higher than the average performance (otherwise, one
cause could be a mismatch between the reference set and what is
discovered). We only keep candidate and reference groups that are size
3 or higher, consistent with \cite{leskovecNature2018}.

\subsubsection{An Example of Evaluation}
\label{sec:example}
Assume the reference group has two groups, $\rgrp_1$, $\rgrp_2$, with
sizes 10 and 15, while we obtained 3 candidate groups $\cgrp_1$,
$\cgrp_2$, $\cgrp_3$, with sizes respectively, $6, 14, 5$, from a
single run of the community detection algorithm. Say $|\cgrp_1 \cap
\rgrp_1 | = 5$, and $|\cgrp_1 \cap \rgrp_2 | = 1$. Recall of
$\cgrp_1$, when matched to $\rgrp_1$ is $\R(\cgrp_1, \rgrp_1) =
\frac{5}{10}$. The precision from this match is, $\Prec(\cgrp_1,
\rgrp_1) = \frac{5}{6}$. Therefore, $\match(\cgrp_1, \rgrp_1 ) =
\sqrt{\frac{5}{10}\frac{5}{6}}$, and similarly $\match(\cgrp_1,
\rgrp_2 ) = \sqrt{\frac{1}{15}\frac{1}{6}}$, so overlap score of
$\cgrp_1$ is $\gscore(\cgrp_1)=\match(\cgrp_1, \rgrp_1 )\approx 0.65$
Assume $|g_2 \cap \rgrp_2 | = 14$ and $|g_3 \cap \rgrp_1 | = 5$. Then
$\gscore(\cgrp_2)=\match(\cgrp_2, \rgrp_2 ) =
\sqrt{\frac{14}{15}\frac{14}{14}}\approx 0.97$, and
$\gscore(\cgrp_3)=\match(\cgrp_3, \rgrp_1 ) =
\sqrt{\frac{5}{10}\frac{5}{5}}\approx 0.71$ (note: $\rgrp_1$ is the
best match to two of three candidate groups). Thus the sequence of
overlap scores is $[0.65, 0.97, 0.71]$ ($\cgrp_1$ has 0.65, etc.), and
average overlap score of candidate groups (average 'performance') is
$\frac{0.97+0.71+0.65}{3}\approx 0.78$.

Any sequence of scores that yields the reference ranking, \ie assigns its
highest score to $\cgrp_2$ and second highest to $\cgrp_3$, such as
$[0.2, 0.9, 0.3]$, gets a maximum SPR of 1.0 (and top PR of
$0.97$). So $SPR([0.65, 0.97, 0.71], [1, 3, 2]) = 1$, and similarly
$SPR([0.65, 0.97, 0.71], [-1, 10.3, 5]) = 1$.  If a scoring technique
assigns the scores $[2, 4, 5]$ ($\cgrp_1$ gets 2, etc.), then its top
PR is $0.71$, since its top ranked group is $\cgrp_3$ (with overlap
score of $0.71$), and we have $SPR([0.65, 0.97, 0.71], [2, 4, 5]) =
0.5$.  The reverse of reference ranking yields the minimum SPR of -1, \eg
$SPR([0.65, 0.97, 0.71], [3, 0, 1])=-1$.  There can be ties in scores,
so $SPR([0.65, 0.97, 0.71], [3, 5, 3])=\mytilde0.87$. However, if
either scoring function (the overlap scores or a scoring technique)
assigns the same score to all the generated groups, we report
undefined SPR.

In each trial, we can get a different set of candidate groups and
results, as Louvain is randomized and in synthetic experiments a new
graph (and reference groups) is generated. We average all the results over
the trials (results such as the number of groups generated, average
group performance, SPR of each scoring technique, $\cdots$).

\subsection{Experiments on Synthesized Data}
\label{sec:syn}

In these experiments, the graph is built out of 10 groups, each of
size 30.  The {\em internal (edge) probability}, the probability that
an edge exists between a pair of nodes in the group is set at 0.6 for
5 groups and 0.2 for the other 5 groups, following
\cite{leskovecNature2018}. We refer to this graph generation setting
as {\em Syn1}. Two other synthetic settings are explored in the
appendix.

There are also outgoing or 'noise' edges, connecting nodes from
different groups. The {\em edge-noise} (or simply noise) probability
is the probability that an edge exists between a pair of nodes from
different (planted) groups.  The higher the noise probability (when
generating the noise edges), the harder for any discovery algorithm to
fully recover the planted (intended) groups. We vary the noise
probability and plot SPR and top PR (overlap score of top ranked
candidate group) of various ranking methods.


For each possible value of the noise (edge-noise) probability
($[0.025, 0.05, \cdots, 0.225]$), we average the results over 200
trials, where in each trial a new graph is constructed and Louvain is
run to generate the groups, and each scoring method is evaluated. Note
that as we increase the noise, overall graph edges increase, and the
groups with lower internal density are more prone to 'corruption'
(their nodes may join other groups).



\begin{figure}[!htbp]
\begin{center}
  \centering
  \subfloat[Spearman rank correlations (SPR) with reference ranking, of various ranking methods.]{{\includegraphics[height=9cm,width=8cm]
      {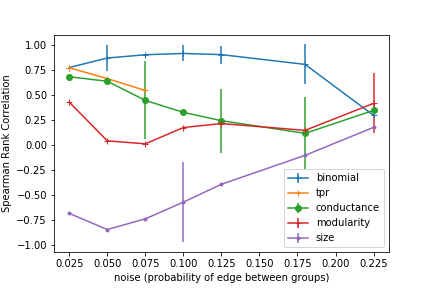} }}
  \subfloat[Mean performance (overlap score) of top ranked group (Top PR). Average of all candidate groups'
    overlap scores in each trial (averaged over all trials) is also shown (avg).]
         {{\includegraphics[height=9cm,width=8cm]{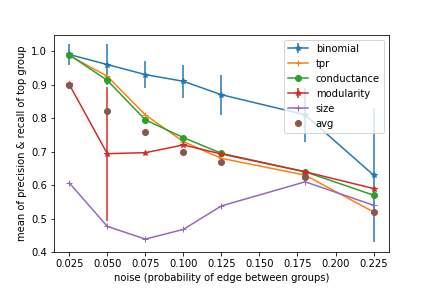} }}%
\end{center}
\vspace{.2cm}
\caption{Planted partition synthesized data, performance averaged over
  200 trials: in each trial a graph is generated, then Louvain is
  applied to generate groups. A sampling of the error bars is shown
  too. As the noise probability is increased (x-axis), the graph
  becomes denser, and fewer (and larger) groups are generated
  (precision goes down). Average number of groups generated goes from
  $8.1$ at 0.025 edge-noise, down to $6.3$ at 0.225. The average of
  groups' overlap scores (avg), shown in (b), goes down as expected
  with increasing edge-noise. }
\label{fig:syns} 
\end{figure}

We observe from Figure \ref{fig:syns} that for a wide range of the
noise probability, the binomial score remains superior based on both
Spearman rank correlation (SPR) and the performance of the top ranked
group. Eventually, the noise is sufficiently high, here at 0.225, that
each method's performance nears that of the average overlap score of
all the groups generated.  At noise of 0.225, the binomial score, in
35 of 200 trials, assigns 0 score (all groups tied), and does not
report a ranking. For these, we score 0 correlation (thus the average
SPR degraded to ~0.25 for Binomial at 0.225 noise). Still, the
Binomial's top group score surpasses others', though, with a very high
error-bar (standard deviation), this is no longer significant.

When noise (probability) is 0 (not shown) or very low, in most trials
there is no ideal ranking in the sense that the planted groups are
perfectly discovered (and all are tied in their perfect overlap
score). So SPR becomes undefined. Similarly for the case of TPR, once
noise-probability goes above $0.075$ the graph and the few groups
discovered become sufficiently dense that all generated groups reach
TPR of 1.0, and no ranking is obtained from TPR.  Observe also that
the plot for Modularity has a similar shape to plot for Size (ranking by descending group
size), although it outperforms it. We will see the high correlation of
Modularity with group size on the real-world datasets as well.

Since in our implementation the approximate node-based binomial score
$\bsn()$ (equation \ref{eq:bn}) is used when $\deg(\g) >= 50$
(equation \ref{eq:bn}), in these experiments the approximate binomial
score is always invoked, as the internal and outgoing edges of the
groups exceed 50. For instance, for a planted group with the lower
density of 0.2, just the expected number of internal edges, $\din$, is
above 80: $0.2 (\frac{30(30-1)}{2}) > 80$.  Indeed, the average
binomial score $\bsn()$ of the top ranked group when noise is low at
0.025 is over 300, consistent with the discussion of Section
\ref{sec:range}, while at 0.225, it goes down to 2.5 (0 $\bsn$ in many
trials).

The Syn1 set up follows that of \cite{leskovecNature2018}, and the
authors report the SPR of a few ranking techniques, in particular
Conductance and Modularity rankings, in their Figure 2. However, we
could not locate the noise rate(s) used for the Figure, in their
publication. From their Figure 2, we note that when Modularity and
Conductance are near respectively 0.3 and 0.6 SPRs (simultaneously),
in Figure \ref{fig:syns}(a)) (a noise of \mytilde0.03), Binomial
ranking scores near 0.8, which is a similar performance to the
performance of CRank reported in their Figure 2. We note that CRank is
an aggregation technique, combining multiple rankers/scorers. It is
promising that a single scoring technique appears competitive with it.



In Appendix \ref{app:apps}, experiments on two other planted partition
settings, equal group sizes and equal probabilities (Syn2) as well as
unequal group sizes (Syn3), are reported with similar patterns of
findings. There, we also observe that with a sufficiently high noise,
Binomial per our approximation implementation returns 0 scores, since
the approximation to tail probability reaches or exceeds the maximum
of 1.0 (thus yielding no ranking, similar to TPR), while Modularity
and conductance keep some discriminatory power.  Some of those groups
are indeed insignificant, and using the exact formula as a special
fall-back case (or the normal approximation) may help in those cases.  On
the real-world datasets, we find that there are ample candidate groups
generated with very high significance, and this significance feature,
of warning the user on certain groups with little or no significance,
will be useful.


\subsection{Experiments on Real-World Data}
\label{sec:real}

We now report on several real-world networks, where inherent
complexities include the widely varying number of connections and
connection patterns across nodes, and highly varied community sizes,
overlapping communities, and complex community hierarchies spanning
several scales.  These complexities are not easily modeled in
synthetic experiments, since the parameters are not known and can vary
substantially from one problem or domain to another.  On the other
hand, some of these attributes may actually, in certain cases,
simplify the task of group discovery or group ranking.

\begin{table}[!htbp]
\centering\small
\begin{tabular}{|c|c|c|c|c|}    
  \hline
  {\bf Dataset}          & {\bf nodes} & {\bf edges} & {\bf reference groups} & {\bf reference group sizes} \\\hline
  Lazega Lawyers \cite{lazega2001,gt2017} & 70 & 378 & 3 & 4, 19, 48 \\\hline
  Football \cite{girvan2002} & 115 & 613 & 12 & 5, 7, $\cdots$, 12, 13 \\\hline
  Railway \cite{gosh2011} & 301  & 1224 & 21 & [1, 1, 4, $…$, 34, 46] median is 13   \\\hline
  Political books \cite{polbooks} & 105 & 441 & 3 &  13, 43, 49  \\\hline
  Political blogs \cite{polblogs} & 1224 & 16716 & 2 & 732, 758  \\\hline
  EU-core \cite{snapnets} & 986 & 16064 & 42 & [1, 1, 2, 3, $…$,  92, 109] median is 14 \\\hline
  YouTube (text) \cite{indep2013} & 7k & 30k to 1mil & 30 & roughly uniform sizes  \\\hline
  20 newsgroups (text) \cite{newsgrps20}  & 7.5k & 70k to 800k & 20 & roughly uniform sizes \\\hline
  Coauthorship \cite{permanence2014} & 90k & 187k & 24 & [29, 154, $…$, 11035, 12674] median is 2058 \\\hline
  YouTube (Social) \cite{snapnets}  & 90k & 1.2mil & 5.7k & [3, $\cdots$, 1601, 1961]  median size is 5  \\\hline
  LiveJournal (Social) \cite{snapnets} & 150k & 6mil & 98k  & [3, $...$, 6k, 47k] median size is 3 \\\hline
  Amazon (Purchases) \cite{snapnets}  & 335k & 926k & 75k & [3, $\cdots$, 29k, 44k, 53k, 54k] median is 5  \\\hline
\end{tabular}
\caption{12 real-world datasets, ordered roughly by size, with number of nodes, edges, and
  reference groups, and summary information on the distribution of the
  reference group sizes.  }\label{tab:datasets} 
\end{table}





We present and discuss the results on the smaller datasets first
(Table \ref{tab:small}), then move to the two text clustering ones,
and finally on the remaining big datasets.  In these datasets, unlike
the experiments on the synthesized graphs, the only source of
variation from one trial to the next is the randomized Louvain
algorithm, which is fairly stable.  Consequently, we observe lower
variance in the results.


\subsubsection{Smaller Datasets}
\label{sec:small}

\begin{table}[htbp]
\centering
\begin{tabular}{|c|c|c|c|}    
\hline
{\bf Method} & {\bf Sp. Rank Corr.} & {\bf Top PR} & {\bf Sp. Corr. With Size} \\\hline
\multicolumn{4}{|c|}{ {\bf Lawyers:} 3 ref. groups, 3 groups generated on avg, \avgPR=0.74  }  \\\hline
Binomial  &  {\bff 0.99}$\pm 0.07$ & {\bff 0.96}$\pm 0.04$  & -0.5   \\\hline
TPR  & 0.03 &  0.76 & -0.82  \\\hline
Conductance  & 0.56  & 0.75  & 0.32   \\\hline
\modularityr &  0.45  & 0.72 & 0.49  \\\hline
(Descending) Size &  0.51 & 0.71  & 1.0 \\\hline
\multicolumn{4}{|c|}{ {\bf Football:} 12 ref. groups, 10 groups generated on avg, \avgPR=0.91 }  \\\hline
Binomial   & 0.36$\pm 0.29$ & 0.84$\pm 0.15$ & 0.2  \\\hline
TPR  & - &  -  & -  \\\hline
Conductance  & 0.34$\pm 0.29$  & 0.84$\pm 0.15$  & 0.4  \\\hline
\modularityr &  -0.13  & 0.77 & 0.91  \\\hline
Size  &  -0.33 &   0.77 & 1.0 \\\hline
\multicolumn{4}{|c|}{ {\bf Railway:} 21 ref. groups, 14 groups generated on avg, \avgPR=0.53  }  \\\hline
Binomial   & 0.41$\pm 0.1$ & 0.61$\pm 0.07$ & 0.93  \\\hline
TPR  & 0.34$\pm 0.09$ &  0.56 & 0.27  \\\hline
Conductance  & 0.12  & 0.48 & -0.67  \\\hline
\modularityr &  0.32  & 0.57 & 0.98  \\\hline
Size  &  0.29 &   0.59 & 1.0 \\\hline
\multicolumn{4}{|c|}{ {\bf Political Books:} 3 ref. groups, 5 groups generated on avg, \avgPR=0.56 } \\\hline
Binomial   & 0.82$\pm 0.11$ & {\bff 0.91}$\pm 0.01$ &  0.84 \\\hline
TPR  & -0.1  & 0.88  & -0.18  \\\hline
Conductance  & 0.81$\pm 0.11$  & {\bff 0.91}$\pm 0.01$   & 0.84  \\\hline
\modularityr & 0.86$\pm 0.09$  & 0.87 & 0.92$\pm 0.01$  \\\hline
Size & 0.78  & 0.89   & 1.0 \\\hline
\multicolumn{4}{|c|}{ {\bf Political Blogs:} 2 ref. groups, 6 groups generated on avg, \avgPR=0.29 } \\\hline
Binomial   & 0.92$\pm 0.05$ & 0.8$\pm 0.004$ &  0.75 \\\hline
TPR  & 0.82  & {\bff 0.88}  & 0.82  \\\hline
Conductance  & 0.58  & {\bff 0.88}   & 0.58  \\\hline
\modularityr & {\bff 0.97}$\pm 0.03$  & 0.87 & 0.99  \\\hline
Size & 0.93  & {\bff 0.88}$\pm 0.001$   & 1.0 \\\hline
\multicolumn{4}{|c|}{ {\bf EU-core:} 42 ref. groups, 7.5 groups generated on avg, \avgPR=0.61 } \\\hline
Binomial   &  0.32$\pm0.15$  & {\bff 0.94}$\pm0.06$ & -0.25 \\\hline
TPR  & 0.24  & 0.71 & -0.44  \\\hline
Conductance & 0.09  &  {\bff 0.94}  & 0.1   \\\hline
\modularityr & -0.64 & 0.34 & 0.82  \\\hline
Size & -0.64 & 0.31 & 1.0 \\\hline
\end{tabular}
\vspace*{.2cm}
\caption{Results averaged over 200 trials.  Number of groups
  generated by Louvain (size $\ge 3$) in a trial on average and the average overlap score
  of these groups, \avgPR, are also reported. }
\label{tab:small}
\end{table}

A description of the the smaller datasets follows, and results shown
in Table \ref{tab:small}. The reader can think for themselves,
apriori, the extent to which the communities derived from the graph
would be consistent with the reference groups, given the descriptions
and information such as the sizes of the reference groups.  The tables
include the average performance (overlap scores) of the generated
groups (for those with size 3 or higher).

{\bf Lazega lawyers: } Nodes are lawyers, the network is the cowork
relation: an edge exists if two lawyers have worked together on the
same case. The reference groups are the 3 office locations (3 cities)
\cite{lazega2001,gt2017}.  This is the smallest dataset where Louvain
creates the fewest, 3 groups.

{\bf Football: } Nodes are college teams, edges are games, and the
reference groups are teams in the same conference (a geographic area,
such as ‘Atlantic Coast’, ‘Mountain West’,...). Teams play with more
teams in the same conference \cite{girvan2002}.

{\bf Political blogs:} nodes are blogs, in 2004, and edges are
weblinks to other blogs (all crawled). There are 2 reference groups:
conservative and liberal (some are manually labeled, some derived from
a directory) \cite{polblogs}.

{\bf Political books: } nodes are books on US politics and edges are
frequent co-purchasing of books (on Amazon). The 3 reference groups
are: neutral, liberal, and conservative \cite{polbooks}.

{\bf Railway: } Nodes are train stations in India, edges connect if
there is some train route with stops on both stations. Groups are
stations in the same state \cite{gosh2011}.

{\bf Eu-core: } Nodes are users (personnel) within a European (EU)
organization and edges are emails among users (an edge corresponds to
at least one email between nodes u and v).  Groups are departments
the users belong to \cite{snapnets}.


Looking at SPR and overlap score of highest ranked, suggest that
Binomial ranking is most consistent with the reference groups.
Football is the dataset where the the generated candidate groups match
the reference groups best (average overlap of 0.91). Interestingly, it
is also the only dataset where the top ranked group score (\topPR) of
no method exceeds the average candidate group overlap score. On this
data, the TPR score was perfect at 1.0, so no ranking was obtained by
TPR (the same issue observed for many synthesized settings, see
Section \ref{sec:syn}).

We observe that the correlation of Binomial with group size is
somewhat mixed, while Modularity is almost always highly correlated
with it, but at the expense of less consistency with the reference
ranking.

\co{
  
  Spearmanr (win if strictly bigger, irrespective of std!) (for
  each dataset one case, default is picked):
  
Binomial vs Modularity: 6, 2, 4 (wins, losses, ties)
Binomial vs Conductance: 9, 2, 1 (win, lose, tie)
Binomial vs TPR: 10, 0, 2 (win, lose, tie)

topf:

Binomial vs Modularity: 8, 2, 2 (win, lose, tie)
Binomial vs Conductance: 6, 3, 3  (win, lose, tie)
Binomial vs TPR: 6, 6, 0  (win, lose, tie)

(if we use top5f instead of topf, when that's available!)
Binomial vs TPR: 7, 5, 0  (win, lose, tie)

Binomial vs Avg: 8, 0, 4 (win, lose, tie)

Conductance vs avg: 6, 0, 6
TPR vs avg: 8, 0, 4

}

\subsubsection{Text Clustering}
\label{sec:clustering}

\begin{table}[htbp]
\centering
\begin{tabular}{|c|c|c|c|c|}  
\hline
{\bf YouTube (text)} & {\bf Sp. Rank Corr.} & {\bf Top PR} & {\bf Top 5 PR} & {\bf Sp. Corr. With Size} \\\hline
\multicolumn{5}{|c|}{  30 ref. groups, edge threshold of 0.05,  26 groups generated on avg, \avgPR=0.76 }  \\\hline
Binomial  & {\bff 0.33} $\pm$0.015  & 0.59 &  {\bff 0.82}$\pm$0.001 & 0.65 \\\hline
TPR  & 0.1  & {\bff 0.7} $\pm$0.001 & 0.68 $\pm$0.001 & -0.25  \\\hline
Conductance & -0.3 & 0.02 & 0.53 & -0.5  \\\hline
\modularityr & 0.11 & 0.56 & 0.72$\pm$0.001 & 0.9  \\\hline
Size & -0.03 & 0.46 & 0.61 & 1.0 \\\hline
\multicolumn{5}{|c|}{  30 ref. groups, edge threshold of 0.1 (110k edges), 28 groups generated on avg, \avgPR=0.79 }  \\\hline
Binomial  & {\bff 0.45}$\pm$0.01 & {\bff 0.95} $\pm 0.001$ & {\bff 0.82}$\pm 0.001$ & 0.59 \\\hline
TPR  & 0.13 & 0.57 & 0.68 & -0.14  \\\hline
Conductance & -0.1 & 0.02 & 0.53 & -0.47  \\\hline
\modularityr & 0.36 & 0.61 & 0.74 & 0.73  \\\hline
Size & 0.06 & 0.54 & 0.68 & 1.0 \\\hline
\multicolumn{5}{|c|}{  30 ref. groups, edge threshold of 0.3 (35k edges), 83 groups generated on avg, \avgPR=0.25 }  \\\hline
Binomial  & 0.85 $\pm$0.004 & {\bff 0.87} $\pm 0.001$ & {\bff 0.82} & 0.96 \\\hline
TPR  & -0.16 & 0.04 & 0.1 & -0.16  \\\hline
Conductance & -0.72 & 0.03 & 0.08 & -0.79  \\\hline
\modularityr & {\bff 0.86} $\pm 0.004$ & {\bff 0.87} & 0.73 & 0.97  \\\hline
Size & {\bff 0.86} $\pm 0.004$ & 0.473 & 0.69 & 1.0 \\\hline
\end{tabular}
\vspace*{.2cm}
\caption{Text clustering on YouTube text (tags) data. 10 trials.  }
\label{tab:utclustering}
\end{table}

\begin{table}[htbp]
\centering
\begin{tabular}{|c|c|c|c|c|}  
\hline
{\bf 20 Newsgroups (text)} & {\bf Sp. Rank Corr.} & {\bf Top PR} & {\bf Top 5 PR} & {\bf Sp. Corr. With Size} \\\hline
\multicolumn{5}{|c|}{  20 ref. groups, edge threshold of 0.05 (772k edges),  12 groups generated on avg, \avgPR=0.41 }  \\\hline
Binomial  & 0.51 $\pm$0.1 & 0.37 $\pm$0.05  &  {\bff 0.52} $\pm$0.05 & 0.87  \\\hline
TPR  & -0.18 $\pm$0.3 & 0.39 $\pm$0.04  & 0.41 $\pm$0.03   & -0.02 \\\hline
Conductance & 0.48 $\pm$0.1 & 0.29  & 0.51 & 0.90  \\\hline
\modularityr & 0.48 & 0.29 & 0.51 $\pm$0.02  & 0.95  \\\hline
Size & 0.46 $\pm$0.1 & 0.24  & 0.50 & 1.0 \\\hline
\multicolumn{5}{|c|}{  20 ref. groups, edge threshold of 0.1 (66k edges),  37 groups generated on avg, \avgPR=0.33  }  \\\hline
Binomial  & 0.87 $\pm$0.01  & {\bff 0.67} $\pm$0.02 & 0.65 & 0.89 \\\hline
TPR  & 0.03 & 0.07 & 0.28 & -0.13  \\\hline
Conductance & 0.1 & 0.02 & 0.24 & -0.08  \\\hline
\modularityr & {\bff 0.89} $\pm$0.01  & 0.65 $\pm$0.04 & {\bff 0.69} $\pm$0.02 & 0.95  \\\hline
Size & 0.86 & 0.40 & 0.61 & 1.0  \\\hline
\end{tabular}
\vspace*{.2cm}
\caption{Text clustering on 20 newsgroups. 10 trials.  }
\label{tab:nwsclustering}
\end{table}

In the clustering experiments, documents are converted to tfidf
vectors and cosine similarity between all pairs of document vectors is
computed \cite{ir2008}.  The pairs with similarity above a threshold
are kept as edges (unweighted edges), and Louvain is run on the
resulting unweighted undirected graph. We explored and show results
for a few thresholds on similarity.  We use two datasets for
clustering: 1) the test partition of the 20 newsgroups with 7.5k
documents, where the 20 classes (20 newsgroup channels) are the
reference groups \cite{newsgrps20}, and 2) A subset of publicly available
collection of vectors from YouTube video user tags from the gaming
genre, with 30 classes (most popular games as class labels)
\cite{indep2013}.

As we increase the similarity threshold starting from 0.05, the
resulting graphs become sparser and more groups and smaller groups are
generated by Louvain (Tables \ref{tab:utclustering} and
\ref{tab:nwsclustering}). We report the \avgPR \ of top 5 groups too,
since the number of groups generated is in the 10s now.

Note that the overall performance of clustering as measured by average
overlap score, \avgPR, of the groups generated may first increase but
eventually begins to decrease, as we increase the similarity threshold. As shown
in Table \ref{tab:utclustering}, the performances of Size and Modularity
improve and may eventually surpass others.  This is because the larger
groups found tend to have highest recall, and their precisions are
competitive with smaller generated groups. A similar trend is seen in
Table \ref{tab:nwsclustering} for newsgroups.

\subsubsection{Coauthorship}

In the Coauthorhsip dataset, nodes are authors, edges are coauthorship
(iff they coauthored a paper together), and the reference groups are
the various fields of publication (subject areas/topics)
\cite{permanence2014}. Each author is assigned to the field they have
most papers in, which yields the reference groups.  In this
dataset, unlike the remaining upcoming large datasets, we have
relatively few, 24, and mostly large reference groups (1000s of nodes
in size), while Louvain generates relatively many groups: nearly 5000
groups (and without a threshold of 3 on size, we'd get 10k
groups!). Especially compared to the upcoming large datasets,
coauthorship is sparse, and the large number of candidate groups is
plausible.  Many such groups are tiny and get a low recall since they
are relatively small (Table \ref{tab:res_author}). These small groups
tend to be ranked high by conductance or TPR. Techniques that are
biased towards larger sizes do well on this dataset (\ie attain
sufficient recall).

We also report the size of top group picked by each method (averaged
over the trials).  We see that TPR and conductance gravitate towards
very small groups.  We also note that over 3k and 4k groups (out of
the 5k generated) tie in their perfect score respectively for TPR and
conductance (\avgPR is averaged over the groups tied at top).

When we increase the minimum group size threshold to 50, we obtain just
over 100 such groups on average from Louvain. The size of the top
group picked by the two methods also increases, and so does the
average overlap score \avgPR (from $0.003$ to $0.05$), and the
performance of conductance and TPR when measured in {\bf \topPR} and
{\bf top 5 PR} (see Table \ref{tab:res_author}) improve substantially
on a relative scale.


\begin{table}[htbp]
\centering
\begin{tabular}{|c|c|c|c|c|c|}  
\hline
{\bf Coauthorship} & {\bf Sp. Rank Corr.} & {\bf Top PR} & {\bf Top Size} & {\bf Top 5 PR} & {\bf Sp. Corr. With Size} \\\hline
\multicolumn{6}{|c|}{ 24 ref. groups, 5k groups generated on avg, \avgPR=0.003 }  \\\hline
Binomial  & 0.47 & {\bff 0.41} & 5132.5 & {\bff 0.20} $\pm$0.006 & 0.92 \\\hline
TPR  & -0.05 & 0.0 & 4.08  & 0.00 & -0.2 \\\hline
Conductance & -0.2 & 0.0 & 4.5 & 0.00 & -0.3  \\\hline
\modularityr & 0.47$\pm$0.001 & {\bff 0.41} & 5132.5  & {\bff 0.2}  & 0.94  \\\hline
Size & {\bff 0.48}$\pm$0.001 & {\bff 0.41} & 5132.5  & {\bff 0.2} & 1.0 \\\hline
\multicolumn{6}{|c|}{ 24 ref. groups, 104 groups with min size of 50 generated on avg, \avgPR=0.05 }  \\\hline
Binomial  & 0.64 & {\bff 0.41} & 5132.5 & {\bff 0.20} $\pm$0.006 & 0.94 \\\hline
TPR  & -0.2 & 0.01 &  114.1 & 0.02 & -0.2 \\\hline
Conductance & -0.48 & 0.01 & 70.9 & 0.01 & -0.56  \\\hline
\modularityr & 0.70 & {\bff 0.41} & 5132.5 & {\bff 0.2}  & 0.96  \\\hline
Size & {\bff 0.72}$\pm$0.04 & {\bff 0.41} & 5132.5 & {\bff 0.2} & 1.0 \\\hline
\end{tabular}
\vspace*{.2cm}
\caption{Average of results over 20 trials.  }
\label{tab:res_author}
\end{table}

\subsubsection{YouTube Social}
\label{sec:utube}

Our YouTube social graph is a subset of the original dataset where we
require a minimum degree of 10 \cite{snapnets}, so that  the Louvain
algorithm (Python) finishes in 30 minutes on each trial.  Here, nodes
are users, and edges are friendships. Users can define groups too, and
the reference groups are the connected components in the user defined
groups. There are 5k such reference groups and almost all are very small (Table
\ref{tab:datasets}).

In this dataset, as with remaining large datasets, Louvain creates a
small number of groups (just over 100 here), with widely varying
sizes, as seen in the size of the top picked group by each method in
Table \ref{tab:utsocial}. Binomial's top rank is a group of size 12k
(from output of Louvain) while the largest group (top ranked by Size
and Modularity) is 28k. On the other hand, conductance and TPR pick
sizes 4 and 6 on average. For TPR, many (30) ties at the top with a
perfect TPR of 1.0 (we report their average overlap score for
topPR). Similarly, 12 groups tie in conductance.  No ranking method
does well here in an absolute sense, as judged by \topPR, as the
precision or recall of all groups picked is small (\avgPR is low too,
at 0.1).

In a second set of experiments on this dataset, we restrict the group
sizes to 50 to 500, and take these groups from first pass of Louvain
(\vs the final output), where sufficiently many smaller groups are
generated. This yields \mytilde40 groups on average.  We see that,
with a narrower range of group sizes, overlap performance score
improve for all.



\begin{table}[!htb]
\centering
\begin{tabular}{|c|c|c|c|c|c|}  
\hline
    {\bf YouTube (Social)} & {\bf Sp. Rank Corr.} & {\bf Top PR} & {\bf Top Size} & {\bf Top 5 PR} & {\bf Sp. Corr. With Size} \\\hline
    \multicolumn{6}{|c|}{ 5k ref. groups, 68 groups generated, mean \avgPR=0.1  } \\\hline
Binomial  & {\bff 0.57} $\pm 0.02$  & 0.07 & 12184 & 0.09 & 0.94 \\\hline
TPR  & 0.087 & {\bff 0.17} $\pm 0.00$ & 5.8 & {\bff 0.16} & 0.05 \\\hline
Conductance & -0.28 & 0.06 & 4.4 & 0.04 & -0.3 \\\hline
\modularityr & {\bff 0.57} & 0.08 & 21586 & 0.09  & 0.98  \\\hline
Size & {\bff 0.57} & 0.09$\pm 0.00$ & 27016 & 0.09 & 1.0 \\\hline
\multicolumn{6}{|c|}{ 5k ref. groups, first pass/level of Louvain, 40 groups in [50, 500], mean \avgPR=0.2  } \\\hline
Binomial  & 0.12 & {\bff 0.56} & 339 &  0.24 & 0.83 \\\hline
TPR & {\bff 0.49} & 0.46 & 148 & {\bff 0.32}  $\pm$0.02 & 0.05\\\hline
conductance & 0.07 & {\bff 0.56} & 339 & 0.25 & 0.32\\\hline
\modularityr & 0.08 & {\bff 0.56} & 339 & 0.28 & 0.93 \\\hline
Size & -0.07 & 0.02  & 441  & 0.23 & 1 \\\hline
\end{tabular}
\vspace*{.2cm}
\caption{YouTube social graph, results averaged over 3 trials.  }\label{tab:utsocial}  
\end{table}

\subsubsection{LiveJournal}
\label{sec:live}

Our LiveJournal social graph is a subset of the original dataset where
we require a minimum degree of 80 \cite{snapnets}, so that Louvain
algorithm finishes in under 2hrs in each trial.  Here, nodes are
users, and edges are friendships. A user can define user groups too,
and the reference groups are the user-defined groups.

Binomial and Modularity rank very large groups that don't do as well
as smaller groups with respect to the small reference groups in this
dataset. However, there are sufficiently many smaller groups here for
TPR and Conductance to do well, and \avgPR is fairly high too
(compared to YouTube Social).  There are tens of groups that tie (are
perfect) at the top for both TPR (\mysim 87) and conductance (\mysim
13).  If we constrain the candidate groups to a smaller set of up to
500 maximum size (\mysim87 candidates remain), we observe the
performances become comparable (Table \ref{tab:live}).


\begin{table}[!htb]
\centering
\begin{tabular}{|c|c|c|c|c|c|}  
\hline
    {\bf LiveJournal (Social)} & {\bf Sp. Rank Corr.} & {\bf Top PR} & {\bf Top Size} & {\bf Top 5 PR} & {\bf Sp. Corr. With Size} \\\hline
    \multicolumn{6}{|c|}{  98k ref. groups, 105 groups generated on avg, \avgPR=0.65 }  \\\hline
Binomial  & -0.3$\pm$0.02  & 0.2$\pm$0.002 & 32k & 0.43 & 0.97 \\\hline
TPR  & 0.3$\pm$0.006  & 0.70$\pm$0.002 & 214 & 0.68 & -0.31  \\\hline
Conductance & {\bff 0.4}$\pm$0.003 & {\bff 0.77} $\pm$0.000 & 25 & {\bff 0.75} $\pm$0.000 & -0.25 \\\hline
\modularityr & -0.3 & 0.2 & 32k & 0.28 & 0.98   \\\hline
Size & -0.4 & 0.14 & 38k & 0.3 & 1.0 \\\hline
\multicolumn{6}{|c|}{  98k ref. groups, 87 groups generated on avg in [3, 500] size range, \avgPR=0.71 }  \\\hline
Binomial  & -0.15 & 0.71 & 380 & {\bff 0.79} $\pm$0.000 & 0.95   \\\hline
TPR  & {\bff 0.27}$\pm$0.005  & {\bff 0.74} $\pm$0.01 & 148 & 0.73$\pm$0.01 & 0.15  \\\hline
Conductance & {\bff 0.28} $\pm$0.01 & 0.71 $\pm$0.000 & 24 & 0.70$\pm$0.000  & -0.08  \\\hline
\modularityr & -0.17  & 0.71 & 380  & 0.74 & 0.97  \\\hline
Size & -0.29 & 0.36 & 415 & 0.47$\pm$0.01 & 1.0 \\\hline
\end{tabular}
\vspace*{.2cm}
\caption{Experiments on LiveJournal, 5 trials.  }\label{tab:live}  
\end{table}

\subsubsection{Amazon}
\label{sec:amaz}

In this dataset, nodes are items (products), edges are commonly
co-purchased pairs of items \cite{snapnets}. The reference groups are
defined by product category: they are connected components of items
with same category. This yield cohesive high precisions groups, but
rather small groups.  Here, the pattern of performance is similar to
the pattern on the LiveJournal graph (Table \ref{tab:amaz}): there are
many small reference groups, and the few in Louvain's output that are
also small match them well, so Conductance and TPR do well.  Binomial
does worse because it scores very large generated groups high, and
such don't find a good enough match in the reference groups.

\begin{table}[!htbp]
\centering
\begin{tabular}{|c|c|c|c|c|c|}  
\hline
    {\bf Amazon (CoPurchases)} & {\bf Sp. Rank Corr.} & {\bf Top PR}
    & {\bf Top Size} & {\bf Top 5 PR} & {\bf Sp. Corr. With Size} \\\hline
    \multicolumn{6}{|c|}{  75k ref. groups, 238$\pm 8$ groups generated on avg, \avgPR=0.54 }  \\\hline
Binomial  & -0.49$\pm0.05$  & 0.54$\pm0.02$ & 11894 & 0.53$\pm0.003$ & 1.0 \\\hline
TPR  & 0.24$\pm0.04$ & {\bff 0.88}$\pm0.05$ & 31.4 & {\bff 0.90}$\pm0.03$ & -0.5  \\\hline
Conductance & {\bff0.74}$\pm0.02$ & {\bff 0.80}$\pm0.00$ & 157 & {\bff 0.83}$\pm0.02$ & -0.7 \\\hline
\modularityr & -0.50$\pm0.03$ & 0.54$\pm0.02$ & 11894 & 0.53$\pm0.003$  & 1.0  \\\hline
Size & -0.51$\pm0.03x$ & 0.54$\pm0.1$ & 12249.0 & 0.51$\pm0.01$  & 1.0  \\\hline
\end{tabular}
\vspace*{.2cm}
\caption{Experiments on Amazon. 5 trials.  }\label{tab:amaz}  
\end{table}

Appendix \ref{app:purity} explores correlation of the ranking
techniques with ranking based on the minimal number of product
categories needed to cover each group.  As in the previous sections,
when the group sizes span multiple scales, binomial (and modularity)
prefer larger groups that require bigger cover sets \vs conductance
preferring smaller groups with smaller covers. However, as we narrow
the range of group sizes scores, we observe the differences shrinks,
and eventually the trend actually reverses.


As the analysis of Section \ref{sec:range} explained, the binomial
scores for bigger groups can be very high. For example, for the Amazon
dataset, the highest scoring generated group gets a binomial score of
over 43k (observed over multiple runs), with group size over 11k
items (size shown in Table \ref{tab:amaz}), with $\din$ over $35k$,
while $\deg$ is $\approx 40k$. Note that this group is indeed
substantially significant: a group with less than 4\% of nodes has
nearly 90\% of its many incident edges as internal.  A few of this
group's most common topics, up to depth 3 in the category hierarchy,
include 'Books\_Subjects\_Children\'sBooks',
'Books\_Subjects\_Nonfiction', and 'FolkTales\&Myths\_Stories',
covering respectively 9.5k, 181, and 153 of in the group. Many of the
other groups also have 'Books' as their top category, but percent
covered by 'Books\_Subjects\_Children\'sBooks' is far below this.

The median binomial score of the 200+ groups generated is over 800,
and the minimum remains over 40 (which is the score of the group with
smallest size 11), thus all groups generated are highly significant in
this dataset. We do see borderline or significance groups in other
cases (\eg see Appendix \ref{app:bin}), but in our experience, there
exist numerous groups generated with healthy significance in
real-world datasets.

\subsection{A Comparison Summary}


\begin{table}[!htbp]
\centering
\begin{tabular}{|c|c|c|c|c|}  
\hline
Binomial \vs $\rightarrow$ & Size & Modularity & Conductance & TPR \\\hline
SPR &  (wins=9, losses=2, ties=1) & (6, 2, 4)  & (9, 2, 1)  & (10, 2, 0) \\\hline
\topPR & (8, 2, 2) & (7, 2, 3) & (6, 3, 3)  & (6, 6, 0) \\\hline
\end{tabular}
\vspace*{.2cm}
\caption{Number of wins, losses and ties of Binomial \vs 4 ranking
  methods, based on Spearman Rank Correlation (SPR), and the overlap
  score of top ranked group, on the 12 datasets. In case of multiple
  experiments (for Newsgroups, CoAuthorship, $\cdots$), we use the
  first/default set up.  A 'win' is whether the (average) performance
  is higher.  }\label{tab:one-on-one}
\end{table}



In the absence of a clear and single source of groundtruth, it is
difficult to make sweeping conclusions.  We aimed to see how
consistent, across a variety of datasets, the rankings by various
techniques were with the reference rankings, obtained via matching
against the groups provided for each datasets. It is very possible
that a group ranked high is informative and useful, but scores low
according to the matching criterion with any of the available
reference groups \cite{gt2017}.  Table \ref{tab:one-on-one} gives a
summary of pair-wise comparisons over all the 12 datasets (using the
default or first setting for each dataset, in case of multiple
experiments). We obsever that, due to the nature of the ranking task
and available reference groups, even a simple baseline such as Size
can score wins.  Overall however, given the results from synthetic
experiments and the often competitive ranking performance of binomial
ranking on many of the diverse datasets, we can conclude that binomial
modeling with its simplicity and efficiency offers a robust
alternative that complements existing scoring techniques well.

Of course, Table \ref{tab:one-on-one} hides the causes behind the wins
or losses (as well as extent of win/loss).  As explained in the
previous subsections, the findings above suggest a pattern: when the
reference groups are few and relatively large, or when the candidate
groups are in a similar size range (\eg within at most an order of
magnitude, ie up to 10x, of each other), we expect Binomial ranking to
do well and often better than Conductance, TPR, or Modularity.  As
seen above, and per our motivation, Binomial tends to score larger
groups, which provide more evidence of community,\footnote{More
  accurately, the larger candidate groups in real-world datasets,
  provide more evidence that they are far from being generated by a
  random process.} higher, although this preference, \ie the
correlation with Size, is not as large as the case for the
(group-wise) Modularity ranking, as observed in the tables.  We also
note that statistical significance of a property is not the same as
strength of a property: given some quantification of the 'community
property', it is possible that a group exhibits large statistically
significant community property, but nevertheless the strength
or intensity of the property can be relatively weak.  For example,
conductance can be one measure of intensity. Another is the ratio of
observed to expected connection proportions, $\i=\frac{\q}{\p}$, as
explained in Section \ref{sec:discuss}. We leave exploring and
utilizing such distinctions to future work.

\section{Related Work}
\label{sec:rel}

The notion of confidence and in particular statistical confidence has
a range of meanings and approaches in network analysis and community
detection.  A comprehensive classification and summarization of these
methods appears in He \etal \cite{he2018detecting}.  In particular
much of the work has been concerned with the significance of the
entire partitioning discovered, and not on single group (candidate community) at a time
\cite{metrics2017,fortunato2016}. Those working on significance
of individual groups, have focused on developing algorithms for group
discovery (community mining) and do not evaluate ranking given groups
\cite{he2018detecting}, with the exception of the work of Zitnik \etal
\cite{leskovecNature2018}, based on which we designed much of our
evaluation protocol and experiments. Their work shows among other
findings that there can be significant room for improved ranking over
random or arbitrary rankings, as well as over simple measures, in particular
Modularity and Conductance. Our work shows that a simple statistics
based measure can also substantially improve over exiting techniques, and
meet our original motivation for a preference towards larger group sizes.



Per classification of He \etal \cite{he2018detecting}, our work falls
under the analytic approach for computing significance of a single
community, but differs from others in that we consider a conditional
event, that given is a group of $\deg$ many incident edges, and our
notion or candidate for an extreme event is observing $\din$ or more
internal edges.  In the existing approaches, a background model such
as ER (Erdos-Reny) random graph model \cite{koyuturk2007,su2018} or
the configuration model \cite{he2018detecting,miyauchi2015} is
assumed, and given a scoring function reflecting a desired community
property, such as density, a formula is derived for probability of
observing a subgraph scoring above the observed value. One challenge
is coming up with a good definition of community property.  For
example, density (such as $\frac{\din}{\deg}$), ignores the total
degree and size of the group. Another challenge is deriving a simple
bound that is sufficiently tight.

The derivation of the binomial formula is simple and short, and the
approximation provides both a lower and upper bound (thus one can
bound the approximation error).\footnote{There may also exist tighter
  bounds, but possibly limited to special cases.} We briefly compare
with the upper bound formula of He \etal \cite{he2018detecting} (in
Section \ref{app:bin}), and find that on synthetic datasets we tested,
node-based binomial has a wider range, and consistently leads to
better ranking. The authors' goal is to derive a community mining
algorithm with the bound, and they show promising results.  Subsequent work
of the authors derives an exact bound based on the ER random graphs,
with improved detection results \cite{he2020}, but this bound may be
inefficient as it requires summing factorials up to number of edges.
Traag \etal use density of all the groups found in the partitioning
(called Significance) \cite{densitySignificance2013} and later develop
the Surprise score for assessing an  entire partitioning, which
is based on the number of edges of all groups in the partitioning
\cite{suprise2015}. Further, they use a binomial tail (although not
referred to as such) with number of trials being the lesser of the
total number of possible internal edges across all groups and number
of graph edges, to asymptotically approximate their measure of
surprise, and present an efficient variant of Louvain algorithm for
Surprise \cite{suprise2015}. The KL divergence for approximations also
arises there.  These lines of work show the promise of scoring based
on a statistical significance approach using binomials, for
discovering candidate communities as well.


Several approaches attempt to sample from the space of random graphs
(ER or configuration space), such as \cite{kojaku2018}, but these
becomes prohibitive for large graphs. Other approaches are somewhat
indirect, and attempt to capture significance in terms of nodes that
appear to least belong to a candidate group (\eg
\cite{plos2011,essc2014}).

The work of Yang and Leskovec \cite{leskovec2012icdm} introduced a
number of datasets with reference groups and assessed the performance
of a variety of scoring techniques, based on several goodness
criteria, using the reference groups.  Later work has cautioned about
treating the reference as groundtruth and basing the evaluations
heavily on comparisons to them \cite{gt2017}.  Community analysis
indeed depends much on the domain and the particular goals of the
task.  While we used the provided reference groups, we contextualized
our comparisons to shed more light on the findings.

The binomial model naturally finds numerous applications in various
analyses of random graphs, in counting and deriving the probability of
finding certain structures (\eg dense subgraphs). However, we are not
aware of this direct use of the binomial tail for scoring significance and ranking
of candidate communities. It is a way of operationalizing what a
community means, in particular under the {\em modern} view of what
makes a good community \cite{fortunato2016}, \ie based on a comparison
of estimates of internal (within group) vs external or overall
probabilities (\eg see Equation \ref{eq:kl}), rather than earlier
definitions based on non-statistical functions of raw counts such as
group degree and conductance.

\co{

the work to
derive confidence are based on a random graph model, most making the
configuration model, which assumes random graph that abide by the
degree sequence of a given graph. Some work are then based on sampling
\cite{}.

related \cite{lancichinetti2010}

datasets/groundtruth  \cite{gt2017}

algo \cite{louvain}

significance: \cite{he2018detecting,he2020}

surveys/overviews: \cite{fortunato2009,fortunato2016}
}

\section{Conclusions}
\label{sec:summary}

We developed community analysis based on the binomial tail.  Our
motivation was to design a principled inexpensive scoring method that was
naturally more biased towards group size compared to existing
techniques such as conductance.  The approach is simple, efficient,
and parameter-free, and we find that it offers a robust method for
confidence assignment and filtering. Extensive experiments
demonstrated the utility of the binomial score for ranking. The
modeling space is rich, and we explored several variants, such as node
and edge-based null models, and an enhancement using the union bound.
We also described how the tail is applicable to two other tasks:
evidence of community membership and assessing edge significance in
the community-induced graph. Given its simplicity and versatility,
scoring based on the binomial tail offers a basic handy tool for
network analysis and beyond.

\section*{Acknowledgments}
Thanks to the machine learning reading group at Tetration, for
discussions as this research progressed, and to the members of
customer adoption and services teams for providing us network data and
feedback.

\bibliographystyle{unsrt}

\bibliography{global}

\appendix

\section{Experiments on Additional Synthetic Settings}
\label{app:syns}

Here, we look at two more distributions for the planted partition
model, in particular to see whether similar patterns from Section
\ref{sec:syn} hold. As observed previously, we find that binomial
ranking is a substantially better ranker over a considerable noise
region, but we also observe limitations of coverage and accuracy of
the approximation to the tail in a high noise region. These issues are
unlikely to be common in practice, based on our experiments on
real-world data, where there often exist ample highly significant
groups.  We will also briefly mention the solution of using exact or
other (\eg normal) approximations in these cases of borderline groups.

\begin{figure}[htbp]
\begin{center}
  \centering
  \subfloat[Spearman rank correlations (SPRs) with ideal ranking]{{\includegraphics[height=8cm,width=9cm]
      {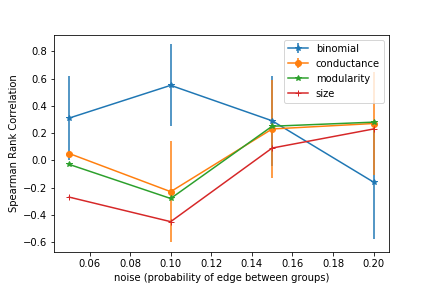}}}
  \subfloat[Mean performance (overlap score) of top ranked group]{{\includegraphics[height=8cm,width=9cm]
      {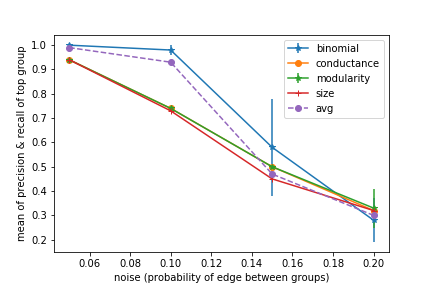}}}
\end{center}
\caption{Planted partition synthesized data, equal group sizes and
  equal group (internal) edge probabilities (all 0.4, Syn2), averaged over
  200 trials. Binomial ranking does a better job until noise becomes sufficiently high.  }
\label{fig:syns_equal} 
\end{figure}

On both the datasets below, TPR always returned 1.0 on the generated
groups, and thus yielded no ranking, and it's not shown.  First, we
look at equal group sizes as well as equal internal edge probabilities
for all the 10 groups, {\em Syn2}: each of size 30 and internal
probability of 0.4, shown in Figure \ref{fig:syns_equal}.

Similar to results of Section \ref{sec:syn}, initially the problem is
easy and then follows a range of noise probabilities where Binomial
ranking separates itself from the rest.  However, as the problem gets
too noisy, Binomial, as implemented based on the approximation to the
tail, loses its dominance, and all the techniques perform near the
average overlap score of the generated groups, which goes down to 0.3
(Figure \ref{fig:syns_equal}(b)).  Initially at 0.05, Louvain
generates $\approx$10 groups in every trial. In almost half the trials
at 0.05 noise, there is no ideal ranking (all groups are perfectly
recovered) and the results are averaged over the remaining half.  With
the edge noise raised eventually to 0.2, the number of generated
groups goes down to near 7 on average.

\begin{figure}[htbp]
\begin{center}
  \centering
\subfloat[Spearman rank correlations with ideal ranking]{{\includegraphics[height=8cm,width=9cm]{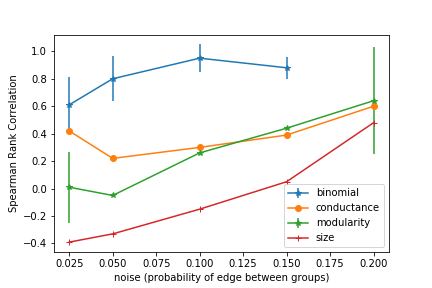} }}
\subfloat[Mean overlap of top ranked group]{{\includegraphics[height=8cm,width=9cm]{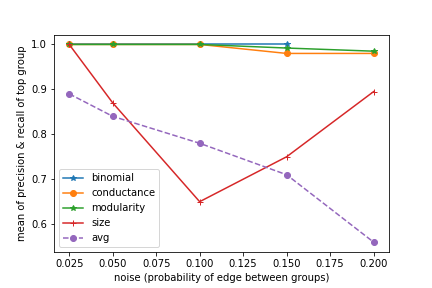} }}
\end{center}
\caption{Planted partition synthesized data, uneven group sizes with equal group edge probabilities (Syn3),
  averaged over 200 trials.  } 
\label{fig:syns_uneven} 
\end{figure}

Next, Figure \ref{fig:syns_uneven} shows a case of 10 groups with
unequal group sizes, {\em Syn3}, $[160, 60, 50, 40, 40, 30, 30, 30,
  30, 20]$, but equal internal probabilities, each of 0.4. Here, we
see a similar pattern. However, in the high noise region of 0.2, in
most trials all groups are tied with 0 $\bsn()$ score under Binomial,
while the Modularity and Conductance, as well as Size, still do well
in selecting a top group (which is often the largest candidate group).
The number of groups generated (in a trial) began with nearly 7 at
0.025 and went down to 4 at 0.20. At noise level of 0.25 (not shown),
performances Modularity, Conductance of go down to 0.46, while avg
group score goes to below 0.3.

In the implementation of binomial scoring, $\bsn()$, if the observed
ratio is not higher than the expected, \ie $\frac{\din}{\deg}\le \p$,
we return 0 score (insignificant). Indeed, looking at the ratio of the
4 groups generated, in all trials, the observed ratio was smaller or
close to the expected, at noise level of 0.2.  If we drop the 0 check,
we often get negative group scores for $\bsn()$ at noise level of 0.2,
and ranking is not improved either.  In this region, while the group
score is statistically insignificant, the ranking task may remain
useful.  We replaced the approximation to the binomial tail with an
exact computation of the tail (Equation \ref{eq:binomial}), at 0.2 noise.
Each trial took a minute or so\footnote{Implemented with several speed ups for the factorial, including
doing the complement when faster.}
(while 200 trials would take well under
a minute with the approximation), but the better binomial scores
(probabilities) indeed led to superior rankings over the other
techniques: For Binomial, Conductance, and Modularity, the SPRs (with
the reference ranking) were respectively 0.630, 0.56, 0.57, and \topPR
values were 0.987, 0.929, 0.987, \ie the largest group was often
picked (averages over the 10 trials).  We looked at the (exact)
binomial score obtained by the top ranked group (group with the
highest binomial score), and indeed it was very poor at $\mytilde0.3$
on average. Note that a score of 1 or higher corresponds to a pvalue
of near 0.1 (when significance becomes noticeable).

We could use the exact form of (or the normal approximation to) the
binomial in such exceptional scenarios (when $\q \approx \p$) to get
better (approximate) scores, a possible improvement for future
work. But the utility of this extension is perhaps very limited in
practice: On the real-world datasets, in our experiments, there have
been ample candidate groups generated that obtain very high
scores. Regarding coverage and significance, see also Table
\ref{tab:variants}. We note however that, in a community mining
context, for a bottom-up search algorithm seeking to improve the
binomial score in each step (as an objective), alternate scores
(modularity, intensity, ...) may be useful initially when the group(s)
being generated are not yet significant.

We have experimented with other variations of synthesized graphs, such
as internal planted probabilities decreasing with graph size (reverse
of Syn3), and we have observed similar results.

\section{Explorations of Variants of the Binomial}
\label{app:bin}

\begin{figure}[htbp]
\begin{center}
  \centering
  \subfloat[Equal sizes, unequal probabilities]{{\includegraphics[height=8cm,width=9cm]
      {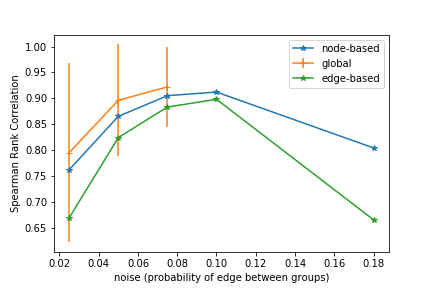}}}
  \subfloat[Unequal sizes and unequal probabilities]{{\includegraphics[height=8cm,width=9cm]
      {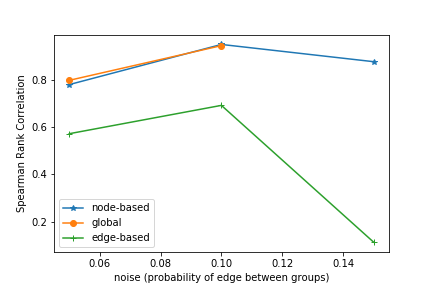}}}
\end{center}
\caption{SPR comparisons of global and edge-based on Syn1 and Syn3
  graph generation schemes.  }
\label{fig:syn_variants} 
\end{figure}

Here, we compare a few variants of binomial scoring to the default
node-based. These are the lower bound score portion of the node-based
scoring, the edge-based model (Section \ref{sec:edgeway}), and global scoring
(Section \ref{sec:global}). Table \ref{tab:variants} and Figure
\ref{fig:syn_variants} show SPR correlations and performances on a few
datasets. We also report on a few aspects such as maximum score
reported and number of groups deemed significant.

We observe that the different variants are very correlated, as might
be expected (Table \ref{tab:variants}).  In particular, the lower bound
score is basically identical in its ranking to the default node-based
(which is the average of high and low), but as we increase the noise,
the lower bound may yield 0 scores sooner than the upper and we
observe the tiny difference in ranking in the high noise region. We
don't report on it further as the differences are tiny.  The global
variant is strict, in the sense that a significant number of groups,
even the reference groups in real-world datasets, are deemed
insignificant or have weak significance, under the global model which
takes the entire graph into account. Some of this is likely due to the
loose upperbound (due to the union bound).

\begin{table}[!htbp]
\centering
\begin{tabular}{|c|c|c|c|}  
\hline
Binomial Variant $\rightarrow$ & Lower Score & Edge-Based ($\bse()$) & Global$^*$ \\\hline
Syn1, 0.05 noise & 1.000 & 0.94 & 0.96  \\\hline
Syn1, 0.1 noise  & 1.000 & 0.94 & --  \\\hline
Syn1, 0.15 noise & 0.998 & 0.93 & --  \\\hline
YouTube Social (5k) & 0.999 & 0.92 & 0.78  \\\hline
YouTube Social (Louvain, 65) & 1.000 & 1.000 & 0.992  \\\hline
Amazon (75k) & 1.000 & 0.999 & 0.896  \\\hline
\end{tabular}
\vspace*{.2cm}
\caption{Correlation (SPR) of default node-based ($\bsn()$) binomial
  scoring and ranking of groups \vs scoring and ranking using three
  variants: Use the lower score (instead of averaging low and high,
  see Equation \ref{eq:approx}), use edge-based binomial score
  (Section \ref{sec:edgeway}), and use the graph adjusted 'global'
  score (Section \ref{sec:global}). The ranked groups are from: groups
  generated on Syn 1 scheme with 3 noise settings (200 trials each),
  reference groups from YouTube Social (5k), output groups from a run
  of Louvain (65 groups generated), and Amazon co-purchase reference
  groups (75k). The correlations are very high.  For the global
  variant, only 300 groups scored above 0 on YouTube, 52 from
  Louvain's output (on several runs), and 25k on Amazon, and the
  correlation is with respect to those subsets of groups. And on Syn1,
  with high noise rates, global returns no ranking on most trials (all
  groups assigned 0), so no SPR is reported on
  such. }\label{tab:variants}
\end{table}

In Figure \ref{fig:syn_variants}, we compare SPR of default node-based
against global and edge-based rankings, in a couple of synthetic
settings, equal size and different probabilities (from Section
\ref{sec:syn}, Syn 1), and different sizes and probabilities (from
Section \ref{app:syns}, Syn 3). We observe again that the global
variant doesn't have the range of the other two (scores become 0 with
raised noise), and edge-based correlates with but appears weaker than
node-based in these two synthetic settings. This underperformance may
be due to how well planted partition models match either node and
edge-based null models.  It is also possible that taking the average
or the maximum of the node- and edge-based scores improves coverage or
ranking performance.  In one experiment on a graph built from hosts
communicating in a datacenter, we observed that the number of groups
with score above a significance threshold of 1.5 went up by
\mysim$5\%$ when using both, over using either one only.  On the
YouTube reference groups, 21 groups get 0 node-based scores and 31 get
below 3, and median score is 11. For edge-based, the numbers are
respectively 24, and over 220, and median score is just 9. Thus
edge-based score seems to be smaller/weaker, but the conservative
global score yields substantially lower scores than both.

We have experimented with the global binomial variant on the
real-world datasets, and we found that, with respect to SPR and
overlap performance of top ranked groups, mostly tied with plain
binomial, and a few mixed results but only small differences. As
observed above, a bigger fraction of candidate groups are deemed
insignificant when using global binomial, but the overall ranking
performances didn't change on the real-world datasets.

\subsection{$p$-value Based on a Configuration Model}

An analytic upper bound of a $p$-value based on the configuration
model (\ie random graphs, same degree sequence) is developed by He
\etal \cite{he2018detecting}, based on the observed density, \ie
internal number of edges $\din$ out of $\deg$ of a group.  The formula
is simple, and we replicate several of the SPR results with
conductance in that paper and find that the approximate binomial bound
appears better for ranking on synthetic experiments, as explained
below. The authors are not focused on ranking in their work, and show
that the community mining algorithm they derive based on the bound is
competitive and promising against other algorithms on a range of of
real-world and synthetic datasets. The $p$-value of a group is upper
bounded by (Equation 4 \cite{he2018detecting}):
\begin{align}
  \label{eq:pvalue}
  \mbox{$p$-value($\g$)} \le \frac{ {\din + \deg  \choose 2\din } {|\edgeset| \choose \din} } {{ 2|\edgeset| \choose 2\din }}
\end{align}

In our experiments, we use the negative log of the $p$-value bound as
we did for the binomial score (rankings won't change), and use
Stirling's approximation to the factorial. Table \ref{tab:pvalue}
shows SPR values, in particular between conductance and the $p$-value
bound on the ranking of reference groups of a few common datasets,
which is match or are consistent with the numbers in the Table 4 of
\cite{he2018detecting}.

\begin{table}[!htbp]
\centering
\begin{tabular}{|c|c|c|c|}  
\hline
  SPRs with $p$-value bound & Football (12) & Railway (21) & Amazon (75k) \\\hline
Conductance  & 0.9231 & 0.95319 & 0.6869 \\\hline
Binomial     & 0.972 & 0.991  & 0.9992  \\\hline  
\end{tabular}  
\vspace*{.2cm}
\caption{Correlation (SPR) of $p$-value upper bound of \ref{eq:pvalue} with
  conductance, and with binomial on the reference groups of 3
  datasets. First row matches He \etal's Table 4
  \cite{he2018detecting} (except they obtain $0.6965$ for $p$-value
  with conductance).  Second row shows high correlation of Binomial
  with $p$-value.}
\label{tab:pvalue}  
\end{table}

\begin{figure}[htbp]
\begin{center}
  \centering
  \subfloat[Syn1]{{\includegraphics[height=6cm,width=6cm]
      {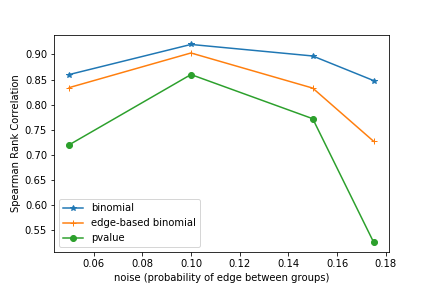}}}
  \subfloat[Syn2]{{\includegraphics[height=6cm,width=6cm]
      {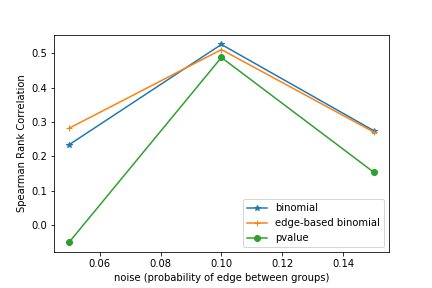}}}
  \subfloat[Syn3]{{\includegraphics[height=6cm,width=6cm]
      {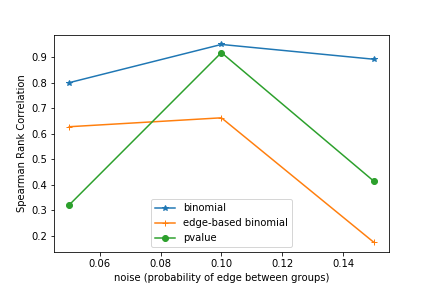}}}
\end{center}
\caption{SPR comparisons of rankings by $p$-value bound of Equation
  \ref{eq:pvalue} with (node-based) binomial and edge-based binomial
  rankings on the three synthetic datasets (200 trials). }
\label{fig:pvalue_sprs} 
\end{figure}

\begin{figure}[htbp]
\begin{center}
  \centering
  \subfloat[Syn1]{{\includegraphics[height=6cm,width=6cm]
      {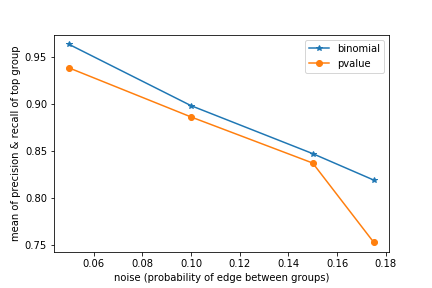}}}
  \subfloat[Syn2]{{\includegraphics[height=6cm,width=6cm]
      {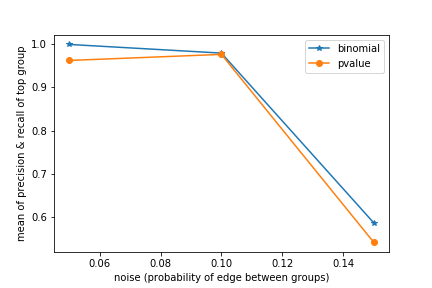}}}
  \subfloat[Syn3]{{\includegraphics[height=6cm,width=6cm]
      {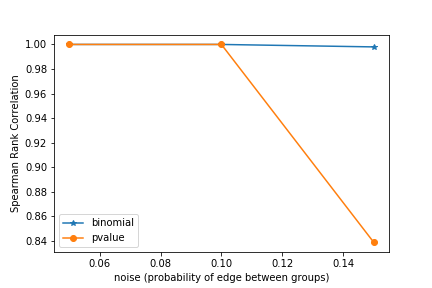}}}
\end{center}
\caption{Overlap score (mean of precision and recall) of top group:
  comparisons of rankings by $p$-value bound of Equation
  \ref{eq:pvalue} with (node-based) binomial ranking on the three
  synthetic datasets (200 trials). }
\label{fig:pvalue_topPRs} 
\end{figure}

We also observe very high correlation of the $p$-value bound with the
(node-based) binomial on these groups. Given that one is based on
group and graph size and the other is based on number of edges, this
is remarkable. However, on the 3 synthetic regimes, the $p$-value
bound was weaker, as shown in Figures \ref{fig:pvalue_sprs} and
\ref{fig:pvalue_topPRs}. One cause for this underperformance is due to
the limited range of the $p$-value bound, akin to the case for our
global binomial. On Syn2 and Syn3 with noise of 0.15, respectively in
around $1/4$th and more than half the trials, all the generated groups
get negative scores, and if we clamp the score to 0, no ranking is
returned. With lower noise rates, the bound underperforms both
node-based binomial score somewhat (with edge-based, it's more mixed),
which may be due to a looser bound, and may not be a significant
difference on real-world data.

\co{

  todo:

  -- correlation of node- and edge-based binomial: show SPR is
  high.. on what groups and data?? synthetic data? we have some data
  already from Amazon

  -- comparison to global binomial (on synthetic, and mention they
  behave similarly on real-world datasets )

  -- min reward (conservative) vs avg for node-based (on synthetic, i
  don't think it's necessary on real-world )

}

\section{Product Category Purity}
\label{app:purity} 

Here, we explore the effect on ranking of the candidate groups having
very different sizes, spanning one or more orders of magnitude, in the
case of the Amazon reference groups.  Furthermore, because the
reference groups are determined in part based on sharing the same
product categories (at lowest level of a product hierarchy), we looked
at correlations with category {\em cover} size, \ie a minimal number
of (product) categories needed to cover a group, where each item in
the group must belong to (be 'covered by') at least one category in
the cover. A smaller cover size may be deemed desirable as it signals
category cohesiveness ('purity'). Conversely, when a group requires a
very large cover, it may be deemed suspect as it signals that items
that don't belong are put together in the group. However, a natural
expectation is that larger groups will require larger covers in
general (and as we observe below), and furthermore the product
categories may not be exhaustive and/or co-purchase patterns on which
the graph is based, can also follow complimentary patterns of purchase
in many cases (products from different categories that tend to
complement, \eg serve the same goal).

We do this analysis on 5000 groups picked based on having high quality
on several dimensions.  Not all items in a group have a category, and
some may have multiple. The zero or more categories assigned appear to
be at the leaf. For example, for a group of 106 items, 13 did not have
any category, and a few categories that covered the most items were:
'books\_Subjects\_Religion\&Spirituality\_Hinduism\_General' (15),
'Books\_Subjects\_Health' (15),
'Books\_Subjects\_Religion\&Spirituality\_NewAge\_General' (8),
$\cdots$.  The groups sizes range from the very small (minimum 3) to
just over 500 items, and the median size is 95. The minimal covers
ranged in size from 1 to 187, with median of 27.

Note that computing the minimum cover is NP-hard, so we applied a
simple randomized greedy algorithm, that repeatedly picks a category
covering most remaining items in the group, until all items in the
group belong to at least one category in the set (cover) picked (note:
items can belong to multiple categories).  Running the algorithm
multiple times gave similar ranking results.\footnote{We also
  experimented with different cover requirement thresholds below
  100\%, \eg 90\%, and saw similar correlation results. }

One might expect the minimal cover size of groups to correlate with
the group size, \ie the more items in a group, the more categories
needed to cover all the items in the group. And this is the case.
Table \ref{tab:amaz2} shows the SPRs, between (minimal) cover size and
a few scoring functions, on the 5000 reference groups where the
conductance, modularity and CRank scores (described in \ref{sec:syn}) are all
provided by the authors \cite{leskovec2012icdm,snapnets} (and we verified
that conductance and modularity matched ours).
We observe that conductance and CRank highly
anticorrelate with cover, while group size (as may be expected),
together with modularity and binomial are highly correlated with it
(column 1 on the left).

Very interestingly, however, as we narrow the range to top 2000 and
top 1000 groups by size, \ie when we compare on more 'local' rankings
(limited size variations), the correlations of all converge to one another:
binomial and to lesser extent modularity move in the direction of
anti-correlation with cover, while this is reversed for conductance.
The last columns show the SPRs as we narrow the range of sizes even
further. In these experiments, in each trial a group is picked at
random (from the 5000), and for a window size of say 100, 100 groups
closest in size, 50 smaller and 50 larger are determined, and the SPR
of the various scores with cover over the 100 groups is recorded. The
SPRs are averaged over 1000 trials.  We see binomial and to a lesser
extent modularity become negatively correlated with cover size,
similar to conductance and CRank, and binomial becomes even more
anti-correlated compared to conductance.  Thus when the group sizes
don't widely vary, we observe that rankings becomes more comparable.

\begin{table}[!htbp]
\centering
\begin{tabular}{|c|c|c|c|c|c|c|}  
\hline
    {\bf SPRs with Cover} & {\bf 5k groups} & {\bf Top 2000} &  {\bf Top 1000}& {\bf Window of 200 }
    & {\bf Window of 100 } & {\bf Window of 50 } \\ \hline
Binomial     & 0.759 & 0.104 & -0.112  &  -0.243 & -0.264 & -0.267 \\\hline
Conductance  & -0.537 & -0.360  & -0.327 & -0.245 & -0.248 & -0.240 \\\hline
\modularityr & 0.807 & 0.223  & -0.024 & -0.195 & -0.223 & -0.237  \\\hline
Size & 0.871 & 0.508 & 0.313  & 0.381 & 0.375 & 0.37 \\\hline
CRank & -0.723 & -0.446 & -0.358 & -0.285 & -0.281 & -0.278   \\\hline
\end{tabular}
\vspace*{.2cm}
\caption{SPRs with a group's product category cover size on 5000
  reference groups. Cover (size) is the minimal number of categories
  that {\em cover} all the products in a group. It is plausible that,
  in general, the larger the group the more categories needed to cover
  it. We observed that size is most correlated with cover on all 5000
  groups (left column), and we observe binomial and modularity which
  correlate with size also correlate with cover, while conductance
  which favors smaller groups, has negative SPR with cover. This is
  also the case with CRank scores, provided with the groups.  However,
  as we narrow the size range of groups compared (top 2000 and 1000
  groups by size, and windows of closest sizes as explained in text)
  we observe that the SPRs converge and furthermore binomial ranking
  becomes substantially anti-correlated with cover, more so than
  conductance.  }\label{tab:amaz2} 
\end{table}

\section{Further Applications of Binomial Modeling}
\label{app:apps}

We first describe an application of binomial modeling to computing the statistical
{\em evidence of group membership}. This membership score, in turn,
could also be used for scoring an entire group, with some benefits as
we describe. We then present an application to discovering significant
edges among groups in the group-induced graph. In all cases, we can
see a pattern: a progression from using plain 'degree' (plain count),
to a ratio (akin to conductance), to binomial modeling (yields a
probability/confidence).

\subsection{Strength of Group Membership}

Often we want to identify node(s) that best represent the group, or
that are most active or connected in the group. For example, we want
to display a few such nodes to the user as a summary so the user more
easily gets the essence of the group. In our application domain, the
nodes are machines and often have names (hostnames), and may have
other attributes (\eg type of OS, ip address(es), $\cdots$), and by
effectively identifying the more central nodes of a group, we may
better summarize or name the group.

One can imagine a variety of measures that reflect different aspects
of importance or belonging (\eg see \cite{centComm2019}).  For
example, one can define a counter part to the group conductance here
by replacing group degree by node degree, and similarly internal
degree becomes the number of edges of the node that are inside the
group.  However, the same drawbacks of normalization remains here for
this measure.  A node that has one or only a few edges, but all
internal, gets perfect score according to this measure.  We can also
easily use the binomial counter part, which in a principled way
combines $\deg, \din$ and $\p$. Note that, like before, $\p$ can be
determined via either using the number of nodes, of the group and the
graph, or the number of edges, of the group and of the graph.

For instance, if we use the node-based null model (using number of
nodes in group \vs graph), with a graph of 20 nodes and a group with 5
nodes, expected probability of connecting to a node inside the group
is $p=\frac{4}{10}$ (when self-arcs are not allowed), and if a node
has $\deg=4$ edges with $\din=3$ inside, its observed connection
probability is $q=\din/\deg=3/4$ and its binomial membership level
(tail score) is around $1.5$ (using the approximations of
\ref{eq:binomial}, $log_{10}$).

Centrality of a node with respect to the whole graph has been studied
\cite{newman2010}, and recently the various centrality measures have
been recently extended to communities \cite{centComm2019}. Degree
centrality (plain node degree) comes closest to the approach here, but
fails to tradeoff external vs internal activity.  The binomial tail
score provides a simple efficient statistical alternative (a
significance value) that takes the tradeoff into account.

Furthermore, one can define a new group score by taking the average or
the median (or some quantile) score of the member scores, each according to a binomial
membership score. Compared to binomial score at the group level (of
Section \ref{sec:form}), the median member score can better reflect
the average 'happiness' of the members and is less prone to preferring
'lopsided' groups as we explain. In particular, it has an advantage
with respect to the resolution-limit problem that a number of other
scores don't (including the group-level binomial score), which we explain
next.

\subsubsection{Group Scores and the Resolution Problem}
\label{app:resolution}

\begin{figure}[!htbp]
\begin{center}
\subfloat         {\includegraphics[width=10cm,height=5cm]{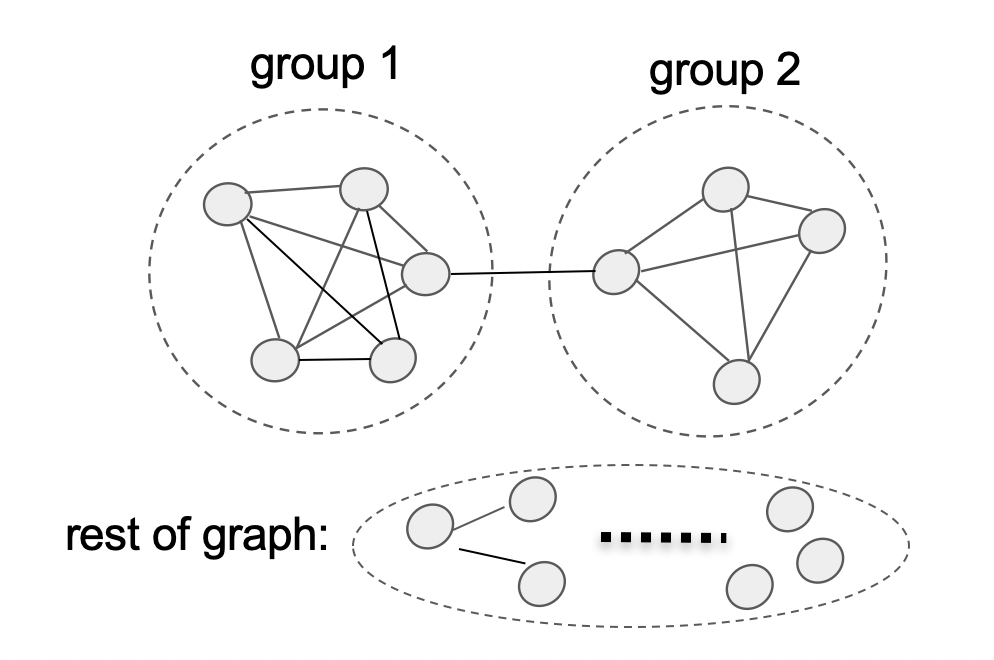}}
\end{center}
\caption{Two dense groups sparsely connected to one another, and
  isolated from the rest of graph.  As long as the rest of the graph
  is sufficiently large, joining the two groups (the union
  ``dumbbell'' group) yields a higher score under several scoring
  techniques, including modularity and group binomial score.  However
  the median of membership scores is higher on each (separated) group.
}
\label{fig:resolution} 
\end{figure}

Consider the two (internally dense) small groups, groups 1 and 2, of
Figure \ref{fig:resolution}, where the two groups are connected via a
single edge only (more generally, relatively few edges) and are
isolated from the rest of the graph.  Intuitively, the two separated
groups are preferred as communities over the single union, dumbbell
group, but if the remainder of the graph is sufficiently large,
modularity scores the union (the dumbbell) higher than groups 1 or
2. This is the so-called {\em resolution limit} problem
\cite{resolution2007,fortunato2009}.  This issue stems from the
phenomenon that the presence of an edge (or a few edges) connecting
the two groups becomes significant as the entire (the rest of) graph
grows: the modularity scores improves when we join the two groups if
the number of edges between them is higher than the expected number of
edges (according to the null model).  The same issue occurs for the (group-wise) binomial scores of
Section \ref{sec:form}. For instance, for the (group-wise) binomial,
in the node-based modeling, letting $p$ be the probability of
connecting for group1, then $2p$ is roughly the probability of
connecting in the dumbbell group, since it has twice the group1's
size. With $d$ internal edges, the binomial tail for group1 is at
least $p^d$, while for the dumbbell group (with at least $2d$ internal
edges) is at most $(2p)^{2d}$, so the dumbbell group obtains a smaller
tail (a higher score), as long as $p^d > (2p)^{2d}$, or $p < 1/2^{d}$
(\eg with $d=5$ internal edges, if group1 is smaller than roughly
$1/2^{5}$ of the entire graph). Interesting, for standard modularity,
if a group is smaller than roughly $\sqrt{|\edgeset|}$ then the resolution
problem may arise.

We can also easily verify similar issues remain for a few other
scores: for instance, for TPR, the scores are tied (no preference on
whether to join or not), while for conductance, joining the two groups
is better, as it yields a perfect conductance score for the union.

Now, considering the membership binomial score for each node, joining
the two groups increases the score for the nodes that have a
connection to a node in the other group (group1 to group2), while
decreases it for the rest (as the group size and thus the probability
of a connection increase, the tail probability under the null model
increases). Thus as long as fewer than half of the group1 nodes have a
connection to the other group, the median score decreases when we
join the two groups.\footnote{Averaging the scores can be influenced
  by extreme scores.  The median of some other node-based score may
  also avoid the resolution problem (\eg membership {\em intensity}
  defined as the ratio of observed probability $q$ of connecting to an
  internal node, to expected $p$).}  Note that we can still join two
cliques that are relatively sparsely connected to one another based on
the node-based measure (as long as half the nodes have an edge to the
other), but in such a situation the union .

Thus, the median membership binomial score can lead to somewhat more
uniformly connected groups (avoids lopsided dumbbells), compared to
the (group-wise) binomial score. Of course, directly optimizing such a
score, for community discovery (community mining), is likely more
challenging as, for example, small local improvements in internal
density may not change the median.  We leave further exploration to
future work.


\subsection{Highlighting Significant Group Interactions}

Edges represent interactions, and once a graph is broken into
candidate communities, an analyst may want to explore the most salient
(significant in some sense) interactions among the discovered groups.
This type of edge analysis can find applications in fields such as
biological/biochemical networks (\eg protein interaction networks) to
draw scientists' attention and effort to the more likely important
interactions. Binomial modeling offer one type of edge (statistical)
significance.


Once groups are discovered, one can build the {\em group-graph} where
groups are the nodes and two groups $\g_1$ and $\g_2$ (the
corresponding vertices in the group-graph) are connected if there
exist any edges among members of the two groups, where the weight of
an edge between groups $\g_1$ and $\g_2$ is the count of such node
pairs.\footnote{The same meta-graph is built after each pass of
  Louvain to improve the modularity objective.}  One may then wish to
first focus on the significant interactions among groups, \ie those
edges with counts more than a null would typically generate.\footnote{A ratio
  measure akin to conductance can be defined, to tradeoff overall
  degree with selectivity to certain group, but that suffers from
  similar aforementioned drawback.}  One null mode can be briefly
described as follows: Given are $k > 2$ groups in the graph. A group
$\g_1$ has $\deg$ total count of (outgoing) edges, and it has $\deg_2$
edge weight to group $\g_2$. We are interested in the tail probability
of observing $\deg_2$ or more connections given the group has several
'choices' (groups) and the choices were picked uniformly at random
among all other group members (thus success probability is $\p =
\frac{|\g2|}{|\nodeset|-|\g_1|}$, where $\nodeset$ is the set of nodes
in original graph).  Note that it's best to use the node size of the
groups to account for different group sizes. There remain several
avenues of exploration.  For instance, as described, an edge between
two groups gets different significance probabilities depending on the
group that we condition on, and we may want to keep this
directionality, or otherwise combine the measures. For the null model,
we also have a choice of whether to account for the count of internal
group edges (group self-arc weights).

\co{ However,

perhaps based on their level

Highlighting nodes that 

naming nodes: when nodes have names, naming a discovered group by
assessing commonality among names of those that are most belong to the
group

similar challenges: normalizing

\subsection{Significance of Edges Among Groups}

}


\section{Node-Proportion Based Modularity}
\label{app:node_louvain}


The modularity objective scores a given set of groups (candidate
communities), and is defined as the sum, over each group $\g$, of the
difference $\delta(\g)$ in observed \vs expected proportion of each
group's internal edges: $\sum_\g \delta(\g)$. As in the
standard applications of modularity, we assume the groups form a
complete partitioning of the nodes, \ie they are exclusive
(non-overlapping) and exhaustive.

The null model in the original paper is based on what we have referred
to as the 'edge-based' model of Section \ref{sec:edgeway}
\cite{newman_2004}, which has the desired property of adjusting for
the chattiness of the group. Here we briefly develop the objective and
the Louvain algorithm based on the node-based model. We are not aware
of this derivation in the literature.

For each group in the partitioning (groups), using the node-based
model, the expected internal edge proportion is
$\frac{\deg}{\m}\frac{\gsize}{\nodes}$. That is, given a group with
$\deg$ incident nodes, the probability that each of the other ends
also falls in the group is $\frac{\gsize}{\nodes}$, thus the expected
number is $\deg\frac{\gsize}{\nodes}$, and the expected proportion is
$\frac{\deg}{\m}\frac{\gsize}{\nodes}$. The observed proportion is
$\frac{\din}{\m}$, and the node-based modularity, denoted $\Qn$, is
therefore:
\begin{align}
\Qn = \sum_{\g \in \mbox{ groups }} \delta_n(\g) = \frac{1}{\m}(\sum_\g \din(\g) - \deg(\g)\frac{\gsize}{\nodes})
\end{align}
We have,
\begin{prop} $\Qn \in [-1, 1]$.
\end{prop} 

We present a proof that $\Qn \in [-1,1]$ (same as edge-based
modularity). First $\Qn\le 1$, as $\frac{1}{\m}\sum_\g \din \le 1$,
and the 2nd term, $\frac{1}{\m} \sum_\g -\deg\frac{\gsize}{\nodes} \le
0$. Next, to show that $\Qn\ge -1$, first note that if the
partitioning consists of one group only, $\Qn = 0$
($\frac{\gsize}{\nodes}=1$ and $\din=\deg$). With two or more groups, we
can replace any internal edge of a group $\g$ with an outgoing edge
that goes from it to another group, and $\Qn$ strictly decreases
(considering both groups affected). Now, with $\din=0$ for all groups,
we note that $\sum_\g \deg(g)\frac{\gsize}{\nodes} \le \m$, as
$\sum_\g \deg(g) = \m$ and $\sum_\g\frac{\gsize}{\nodes} = 1$, and a convex
combination is no larger than the  largest component: when $a_i \in [0,
  1]$, $\sum_i a_i \le 1$, and letting $i^*=\argmax_i b_i$, we have
$\sum_i a_i b_i  \le b_{i^*}$.

For a collection of purely disassortative groups (\ie $\din=0$ for all
groups, such as a pure bipartite graph), we get negative modularity
that approaches -1 as number of nodes grow.  Symmetrically, for many
purely assortative groups ($\forall g, \din(\g)=\deg(g)$), $\Qn$
approaches 1.

Likewise, one could define a simple Louvain algorithm using the $\Qn$
objective. We leave exploring the (practical) differences in the
formulations and, for example, potentials for utilizing both
objectives, to future work (see also Section \ref{sec:two_models}).


\end{document}